\documentstyle[12pt]{article}     
\input epsf		      

\def\abst#1{\begin{minipage}{5.25in}
{\noindent   \normalsize
{\bf Abstract} #1}  \\ \end{minipage}  }
\newtheorem{thm}{Theorem}
\newtheorem{lem}{Lemma}
\def\be{\begin{equation}}
\def\ee{\end{equation}}
\def\bea{\begin{eqnarray}}
\def\eea{\end{eqnarray}}
\def\remark{ \noindent {\bf Remark:}  }
\def\proof{ \noindent {\bf Proof:}  } 
\def\P{Prob_{p_c}}   
\def\E{{\mathbf E}}  
\def\I{\mathrm I}    
\def\C{{\mathcal C}}   
\def\Ltoo{\parbox[t]{.4in}
        {$\longrightarrow \\ {\scriptstyle L \to \infty}$}}
							
\newcommand{\eq}[1]{eq.~(\ref{#1})}	  

\def\lg{\stackrel{\scriptstyle <}{_{_{\scriptstyle >} }} }
\newcounter{masectionnumber}
\newcommand{\masect}[1]{\setcounter{equation}{0}
  \refstepcounter{masectionnumber} \vspace{1truecm plus 1cm} \noindent
    {\large\bf \arabic{masectionnumber}. #1}\par \vspace{.2cm}
      \addcontentsline{toc}{section}{\arabic{masectionnumber}. #1}  
    }
 \renewcommand{\theequation}
    {\mbox{\arabic{masectionnumber}.\arabic{equation}}}
      
    \newcounter{masubsectionnumber}[masectionnumber]
\newcommand{\masubsect}[1]{
    \refstepcounter{masubsectionnumber} \vspace{.5cm} \noindent
  {\large\em \arabic{masectionnumber}.\alph{masubsectionnumber} #1}
    \par\vspace*{.2truecm}
 
\addcontentsline{toc}{subsection}          
 {\arabic{masectionnumber}.\alph{masubsectionnumber}\hspace{.1cm} #1} 
    }
\newcommand{\startappendix}{ \setcounter{masectionnumber}{0} } 
\newcommand{\maappendix}[1]{  
    \setcounter{equation}{0}
  \refstepcounter{masectionnumber} \vspace{1truecm plus 1cm} \noindent
    {\large\bf \Alph{masectionnumber}. #1}\par \vspace{.2cm} 
    \renewcommand{\theequation}
    {\mbox{\Alph{masectionnumber}.\arabic{equation}}}          
   \addcontentsline{toc}{section}{\Alph{masectionnumber}. #1}    
      }                                            

\begin{document}
\begin{figure}[t]{\normalsize    To appear in:   Nucl. Phys. B [FS]} 
\end{figure} 
\title{\vspace*{-.35in}
On the Number of Incipient Spanning Clusters}
\author{Michael Aizenman
\thanks{\protect\normalsize Work supported in part by the 
 NSF Grant PHY-9512729. \protect\newline E-mail: aizenman@princeton.edu} 
\\
\normalsize \it  Departments of Physics and Mathematics  
\vspace*{-0.05truein} \\
\normalsize \it Princeton University, Jadwin Hall 
 \vspace*{-0.05truein} \\
\normalsize \it Princeton,  NJ 08544-0708.}
\date{{\small Sept. 16, 1996; revised Oct. 22, 1996}}
\maketitle
\thispagestyle{empty}        

\abst{
In critical percolation models, in a large cube there will
typically be more than one cluster of comparable diameter.
In 2D, the probability of  $k>>1$ spanning clusters is of the 
order $ e^{-\alpha \ k^{2}}$.  In dimensions $d>6$, 
when $\eta = 0$ the spanning clusters proliferate:
for $L\to \infty$ the spanning probability tends to one,
and there typically are $ \approx L^{d-6}$ spanning 
clusters of size comparable to  $|\C_{max}| \approx  L^4$.  
The rigorous results confirm a generally accepted 
picture for $d>6$, but also correct some misconceptions 
concerning the uniqueness of the dominant cluster.  
We distinguish between two related concepts: the Incipient Infinite 
Cluster, which is unique partly due to its construction, 
and the Incipient Spanning Clusters, which are not.  
The scaling limits of the ISC show interesting differences 
between low ($d=2$) and high dimensions.  In the latter case 
($d>6 ?$) we find indication that the double limit: 
infinite volume and zero lattice spacing, 
when properly defined would exhibit both  percolation at the 
critical state and infinitely many infinite  clusters.  }
\renewcommand{\baselinestretch}{1.5}

\bigskip

\bigskip


critical behavior, hyperscaling, scaling limit.

\newpage
\begin{minipage}[t]{\textwidth}
\tableofcontents
	\end{minipage}
\newpage

\vspace{-1.2cm}
\masect{Introduction} \vspace{-.6cm}
\masubsect{Main Results}

\begin{figure}[ht]
    \begin{center}
    \leavevmode
    \epsfbox{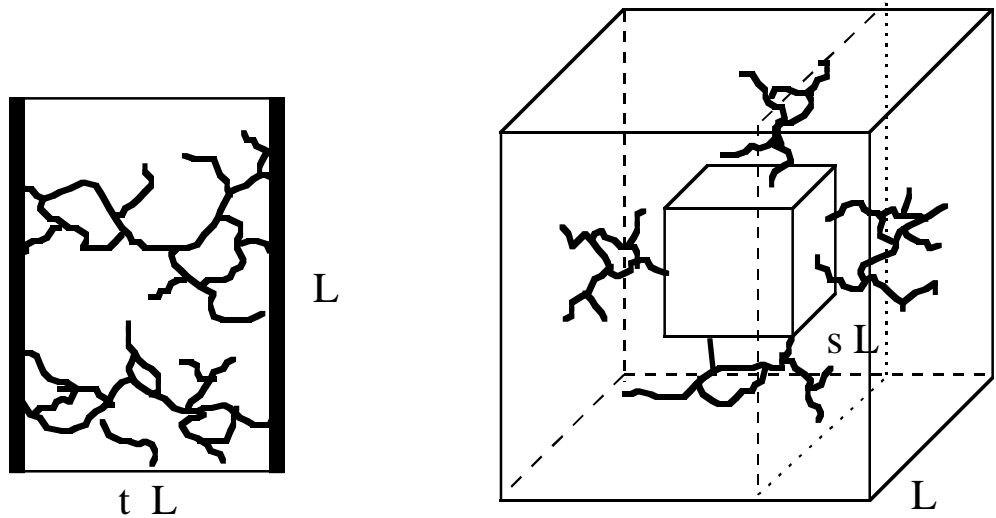}
  \caption{The geometric setups used in the discussion of 
  the Incipient Spanning Clusters.}
\label{fig:cube}
\end{center}
\end{figure}

Some of the essential features of critical phenomena
are reflected in the structure of the large 
correlated clusters observed in the critical state 
\cite{Con,Sta2,SA,Ma}.
For percolation, the relevant notion is of connectivity,
and the characteristic feature of the criticality is the 
spontaneous formation of large connected clusters,
of ``macroscopic'' diameter.  
In this paper we present rigorous results on the number of
Incipient Spanning Clusters (ISC), in various dimensions.
The term {\em Spanning} is applied generically 
to clusters which stretch across a large region, and connect different 
boundary segments (see Figure~\ref{fig:cube},
and the more detailed explanation in 
Section \ref{IIC-ISC}).  {\em Incipient} is the word we use for 
the tenuous structures seen at the critical models, at percolation 
threshold.

Our goals are: i. clarify the issue of uniqueness, concerning which
there has been some confusion,
ii. present new bounds on the distribution of the number of
ISC in 2D models, and iii. prove that in high dimensions 
Incipient Spanning Clusters
behave in a rather different way than they do in 2D.

More explicitly, we establish the following.

\noindent
1) In any dimension $d>1$, at $p=p_c$, the probability
of there being {\em more than one} spanning cluster in a slab of the 
form $[0,tL]\times[0,L]^{d-1}$ does not go to zero, as $L \to \infty$; 
the proof covers the case of $t$ not too large, but independent of $L$ 
(Theorem 2, below).

\noindent
2) For completeness, we add that above the critical point, 
$p>p_c$, there typically is exactly one spanning cluster 
(for $L >>1$) (Theorem \ref{thm6}, Appendix C).  
Below the critical point there is none.

\noindent
3) In $2D$ critical models the number of spanning clusters
is of finite mean, and its probability distribution satisfies:

\begin{equation}
 \P\left(
   \begin{array}{l}
   	\mbox{\small there are exactly $k$ unconnected}  \\
   	\mbox{spanning clusters in $[0,L]^2$}
   \end{array} \right) \
      \begin{array}{ll}
 	   \ge	& A \ e^{-\alpha \ k^2}  \\
 	   \le & e^{-\alpha' \ k^2}
      \end{array} 	
      \label{1.1}
\end{equation}

\noindent 
where $k$ counts clusters connecting the
two opposite faces $\{x_1=0\}$ and $\{x_1=L\}$ of $[0,L]^2$, 
and the bounds hold uniformly in $L$ (Theorem 3).

The power $k^2$ seen in \eq{1.1} represents $k^{d/(d-1)}$.
The analysis suggests  a plausible  argument
for a more general validity of such lower bound in low dimensions,
but it is not clear whether the actual rate is not slower for d=3,4,5. 

\noindent
4)  In dimensions $d>6$, assuming $\eta = 0$  ($\eta$ being the critical 
exponent seen in \eq{eta}):
\begin{itemize}
\item the spanning probability tends to {\em one} (Theorem 4), and
\item
spanning clusters proliferate ---
\begin{equation}
 \P\left(\mbox{the number of spanning clusters $\ge \ o(1) L^{d-6}$}
	 \right) \Ltoo  1  ,	
\end{equation}
in the sense that the limiting statement applies to any substitution 
in which $o(1)$ is replaced by a function of $L$ which vanishes as 
$L\to \infty$.

Furthermore, at 
least for clusters whose size (the number of points) is comparable to 
the maximal ($|\C|_{max}$), $L^{d-6}$ provides the actual rate of 
growth --- not just a bound.

\end{itemize}

\begin{itemize}  
\item The size of the maximal cluster, as well as the size of the maximal
spanning cluster, are typically 
\be |\C^{(sp)}_{max}|\ , \ \ |\C_{max}| \approx L^4 \ , 
\ee 
in a sense 
made explicit in Theorem \ref{thm5}.
\end{itemize}

The assumption on $\eta$ refers to the purported law:
\be
\tau(x,y) \ \equiv \
\P\left( x \mbox{ sites {\em x} and {\em y} are connected} \right)
\approx \frac{Const.}{|x-y|^{d-2+\eta} } \ \ .
\label{eta}
\ee
It is expected that $\eta = 0$ above the upper--critical dimension
 $d=6$ \cite{Tou, HL}.
Rigorous results in this direction were proven by 
Hara and Slade \cite{HS} (reviewed in \cite{Sl}),
who establish that a related, but somewhat weaker condition, holds
at $d>6$ in models with sufficiently spread finite--range connections,
and at somewhat higher dimensions for the nearest--neighbor
percolation model.    

The results proven here for $d>6$ reinforce the generally accepted 
picture, in which the proliferation of large clusters is related to 
the breakdown of hyperscaling (Coniglio \cite{Con}), and 
the ``fractal'' dimension of the large clusters stabilizes at 
$D=4$ (Aharony, Gefen and Kapitulnik \cite{AGK}, 
Alexander et.al. (AGNW) \cite{AGNW}).  Our analysis is based on 
the diagrammatic bounds of Aizenman and Newman \cite{AN}.  The high 
effectiveness of the method points to the validity of the perspective 
offered by AGNW \cite{AGNW} that for $d>6$ large clusters resemble 
randomly branched chains.

The fact that the spanning probability tends to 1
has interesting implications for the
nature of the continuum limit, mentioned in Section~\ref{IIC-ISC}.

Some of the topics discussed here are also the subject of the 
work of C.  Borgs, J.  Chayes, H.  Kesten and J.  Spencer \cite{BCKS},
where the properties of ``incipient infinite clusters'' are
discussed within the context of an axiomatic description of the 
critical behavior.  [The reader is advised, however, that the notions 
have not been coordinated, and terms like IIC and ISC may be used 
in different ways.  The terminology used in this paper is explained 
and motivated in greater detail in Section 5.]

The prototype for the percolation models discussed here, {\em 
and the systems to which we refer by default}, are the 
nearest-neighbor bond, or site, percolation models on the regular 
lattice $Z^d$.  However, with minimal clarifications the discussion 
applies also to percolation models with different short--scale 
characteristics, including finite range bond percolation, n.n.  and 
next--nearest--neighbor site percolation, and also the continuum 
random dots models.  
To attain this robustness, the arguments employed 
refer mostly to large -- scale connection events.  The essential 
features (or at least those assumed throughout) are: independence for 
sufficiently separated regions [a fixed finite distance], and the 
vdB-K property which is presented in Section \ref{sect:2D}.

Following are some less technical comments on the results.

\masubsect{Implications of the Multiplicity of Incipient 
   Spanning Clusters}

The assertion that in 2D there is positive probability for
more than one Incipient Spanning Cluster will not surprise 
mathematicians familiar with the theory developed by Russo \cite{R} 
and Seymour and Welsh \cite{SW}.  Nevertheless, even the 2D case of
the general result (1) is in contradiction with statements 
which  for quite a while have apparently been part of a wide 
consensus. 
That consensus was recently challenged, and 
corrected, by a report of a numerical work \cite{Hu} 
(see also \cite{Ar}).  As we attempt 
to make it plain (and as was already stated in ref. \cite{A_Web}), 
the case of 2D is really simple.  
To prove the result for general dimension, we develop what may be
regarded as a  rigorous real -- space renormalization group argument.

The misconception of the uniqueness of the spanning clusters
could have been facilitated by a number of factors:

\noindent i. {\em Scaling Theory}  \\
The assumption that  there typically is
one dominant cluster (in a finite region $[0,L]^d$)
helps in the explanation of the {\em hyperscaling} law,
which is found to be obeyed in low dimensions.
To re-emphasize that low level multiplicity (up to $L^{o(1)}$)
is consistent with hyperscaling, we present a version of the 
heuristic argument 
in Appendix A.  It is seen there that the relevant condition 
is only that the number of large clusters grows slower 
than any power of $L$.

\noindent  ii. {\em Uniqueness of the Infinite Cluster}  \\
One of the results known under rather general  assumptions
(discreteness and ``regularity'') is that when there is percolation, 
the infinite cluster is unique \cite{NS,AKN,BK} (see however 
Sect.~\ref{IIC-ISC} below).  The Uniqueness Theorems were occasionally 
quoted as an indication that at the percolation threshold, where 
large--scale clusters are expected, there is a single dominant cluster 
--- referred to as the Incipient Infinite Cluster (IIC).  Here, the 
convenient and otherwise appealing terminology might have suggested an 
incorrect deduction.

\noindent  iii. {\em Terminology}  \\
The term Incipient Infinite Cluster is too encompassing.  
The dominant cluster(s) in a large volume can be viewed from two  
perspectives, for which it may be better to have separate terminology.
We propose to keep the term IIC for the large clusters 
as  seen from the perspective of their sites (which presumably
agrees with the conditioned process studied by Kesten  \cite{K3}
in the context of 2D \cite{K3} models). 
The same clusters, as
viewed on the bulk (macroscopic) scale may be referred to as the 
Incipient Spanning Cluster(s).  The distinction becomes more 
clear as one contrasts the microscopic and the macroscopic view of the 
system, and contemplates the relevant mathematical descriptions of the 
{\em scaling limit} --- the limiting situation in which the ratio of 
the two scales diverges.  A more detailed discussion of these issues,
is presented in Section \ref{IIC-ISC}, which can be read in advance of 
the more technical Sections.  
Included in Section \ref{IIC-ISC} are three possible definitions of 
the IIC, which still ought to be proven equivalent. 
For each of them it can be shown that 
if the percolation density vanishes at $p_c$ then  
the IIC process exhibits a unique infinite cluster.  
The ISC in  general are not unique (Section 2).
 
\noindent{\em Red --- Blue bonds} \\
The multiplicity of ISC means also that the important notion of ``red 
bonds'' (Stanley \cite{Sta1, Sta2}, Coniglio \cite{Co82}) needs some 
care.  This notion is best explained in the situation in which a 
macroscopic piece of material is placed between two conducting plates 
which are at different electric potentials, and the bonds represent 
conducting elements.  The ``red bonds'' are often interchangeably 
described as: i.  bonds which are essential for the connection, i.e., 
are unavoidable for every path connecting the opposite boundary 
plates, or as ii.  bonds which carry the full current.  (Other 
elements in this classification are: ``blue'' - the backbone bonds, 
and ``yellow'' - the dangling ends.) As we now appreciate, these two 
definitions do not coincide.

The non-uniqueness of the spanning cluster has the consequence
that frequently the random configuration will not contain any
bond which is essential for the connection (though there will
be bonds essential for their specific spanning clusters).
Interestingly, this does not mean that no bond carries
the full current.  If there are large differences in the
total resistance of the distinct spanning clusters,
than the bulk of the current may still be carried by just one
of them.  Its essential bonds meet the condition
in the second definition of ``red bonds''.
However, as we see now, this terminology involves
more than just the topology of the cluster.

Despite the above complication, it is still desirable to have an 
elementary geometric concept expressing the fact that in typical 
configurations there are bonds whose removal (or blockage) will 
result 
in a change of the available connected routes which is visible 
on a large 
scale $R$.  For this purpose one may define a bond {\em b} to be 
{\em essential on scale $R$} ( $>>$ lattice scale) if there are two 
realized paths, of end--to--end distance $R$ 
emanating from the opposite ends of
the bond, which are not linked by any other path which stays
within the distance $R$ from {\em b}.

Further discussion of the results is found in Section \ref{IIC-ISC}. 


\newpage

\masect{Multiple ISC occur in Critical 
Models in all Dimensions $d> 1$} \label{sect:all_d}

\begin{thm}  (Incipient Spanning Clusters occur)
For any dimension $d>1$, there is an $L$-independent 
(decreasing) function $h_d(t)$,
which is strictly positive at least for $ 0 \le t < 1/2$, 
such that at the critical point for every $0< L< \infty $:
\be
 R_{L,t}\ \equiv 
   \P\left(
	\begin{array}{l}
	 \mbox{ the slab $S_{L,t}\equiv[0,tL]\times[0,L]^{(d-1)}$
	        is traversed}    \\
		\mbox{ in the direction of the 1-st coordinate }
	\end{array} 
         \right) \ \ge \ h_d(t)   \ .
\label{2.1}
\ee
Furthermore, $h_d(\cdot )$ satisfies
\be
h(t) \ \ge \  	1 - A e^{-const. \ t^{-(d-1)}}  \  \ 
 \mbox{ as  } t \searrow 0
\  .
\label{2.2}
\ee
\end{thm}

The two dimensional case of this statement was derived in 
the works of Russo \cite{R} and Seymour and Welsh \cite{SW} 
(conveniently summarized in \cite{G}),
where it is also established that $h_2(t)>0$ for all $0 < t < \infty$.
The result provides a very versatile tool for the 
study of 2D critical models (see below).
We expect that also for arbitrary dimensions $h_d(t)>0$  
for all values of $t$, though that has not been established.
The proof of Theorem 1 is presented in Appendix B.
It uses arguments which may be regarded as standard.

Our main result for arbitrary dimension is the following.

\begin{thm}  (There can be more than one ISC)
For any dimension $d>1$, there is a function $g_d(t)$, 
strictly positive for $0 \le t < t_o$,
 with which for every finite size $L$:
\begin{equation}
   D_{L}(t,p_{c}) \equiv \P\left(
	\begin{array}{l}
  	    \mbox{ there is more than one } \\
	    \mbox{ spanning cluster in $S_{L,t}$}
	\end{array}
	   \right)    \ \ge \ g_d(t)  \ . 
\label{2.3}
\end{equation}
\label{non-uniqueness}
\end{thm}

\noindent {\bf Remarks}:  1)  It should be appreciated that
the bound in \eq{2.3} is satisfied only at $p=p_{c}$.  
Otherwise:
\begin{equation}
	\lim_{L\to \infty} D_{L}(t,p \ [\ne p_{c}]) = 0  \ 
	\label{2.4}
\end{equation}
(for any $t>0$).
For $p<p_{c}$ \eq{2.4} is true because there are no spanning
clusters (the connectivity function decays exponentially), 
and for $p>p_{c}$ it holds because there is only one spanning
cluster (see Appendix C).
 
\noindent 2)  As in Theorem 1, we expect the restriction to small 
$t$, not to be relevant for the strict positivity of $g_d(t)$.
I.e., if the probabilities of the corresponding macroscopic scale 
events are uniformly positive
for some aspect ratio ($t_{o}>0$) then that should also be true
for any other value of $t$.  However, we do not have a proof of
such a principle for $d>2$.  For 2D, the 
strict positivity of $g_2(t)$ is directly implied by the 
RSW theory, as is explained in the next section.

The proofs of Theorem 1 and Theorem 2 (for $d \ge 3$) 
have the flavor of a real - space renormalization argument.
By considering suitably chosen local events, 
determined in blocks of scale $L$, it is shown that for any $p$:

\begin{itemize}
\item[i.]  If $R_{L,t}$ is too small than the connectivity function 
decays exponentially, and hence $p<p_{c}$.  

\item[ii.] If $[1-R_{L/3,3t}] + D_{L,t}$ is too small then there is 
percolation, and $p > p_{c}$.  \\
For $p=p_{c}$ the term $[1-R_{L/3,3t}]$ vanishes as $t\searrow 0$, 
and hence $D_{L,t}$ cannot be too small.
\end{itemize}
The first argument is recapitulated in Appendix B. 
The second is given below.  

\noindent {\bf Proof of Theorem 2:}   As is explained in the 
next section, for 2D the 
RSW theory \cite{R, SW} implies that \eq{2.3} 
holds with a function $g_d(t)$ which is positive for all $t>0$.

To prove Theorem 2 for $d \ge 3$ we shall
show that there is a constant $b <\infty $ 
(based on 2D  considerations, and thus independent of $d$)
such that for every  $p \le p_{c}$ 
\begin{equation}
	(1 - R_{L/3,3t}) + D_{L,t} \ \ge \ b 
	\label{D}
\end{equation}
for any $L < \infty$.  Since $R_{L,t} \ge h_d(t)$, 
that would establish the claim, with
 \begin{equation}
 	g_d(t) =  b - [1- h_d(3t)] 
 \end{equation}
where $\lim_{t\searrow 0} [1-h_d(t)] = 0$, by Theorem 1.

The following argument proves that if (for a suitable choice of 
the constant $b$) \eq{D} fails for some $L$ 
than there is percolation in 
the slab $[0,tL]\times [0,L]^{d-3} \times R^{2}$. 
Let us partition the slab into a  
planar collection of translates
of the $d$-dimensional rectangular cell $[0,tL]\times [0,L]^{d-3} 
\times [0,L/3]^{2}$.  In a given configuration of the percolation 
model, a cell is declared {\em regular} if the following two 
conditions are met:\\
i. the cell contains a spanning cluster (in the first direction), \\
ii.  the block made by joining $3  \times 3 $ cells, 
with the given one at the center,
is traversed by  exactly one spanning cluster.

Notice that if within the 2D array (of d--dimensional regions)
 there is chain of neighboring
{\em regular cells}, then 
the spanning clusters of these cells belong to a common connected
cluster.  

The last observation implies that there is percolation in the slab
unless any finite
region is encircled by a  *-connected loop of cells which are
not {\em regular} in our terminology.  We estimate the probability
of such events by a Peierls--type argument.

While irregularity is not independent for adjacent cell, 
it is independent for cells with a gap of $2$ in one of the two 
direction.  Thus, it is convenient to estimate the probability of the 
existence of an encircling loop by focusing on sub-chains 
of minimal steps with such gap.  
There are $20$ possibilities for each 
step in the chain.  The Peierls estimate, suitably--modified, implies 
that a sufficient condition for percolation of the regular cells is
\begin{equation}
	Prob\left( \mbox{ cell is {\em} irregular } \right) 
	\cdot 	20 \ < \ 1
	\label{Peierls}
\end{equation}

We bound the probability of a cell to be irregular by adding the 
probabilities of the two possible failures.  The result is that there is 
percolation unless \eq{D} holds, with $b = 0.05$.  (That may not 
strike one as a very large number, but still it is positive uniformly 
in $L$).
 
Since the 
probabilities $R_{L,t}$ and $D_{L,t}$ are continuous in $p$ 
(pertaining to regions of fixed finite size), if  \eq{D} fails
at $p$ then it does so also for $p-\epsilon$ for $\epsilon >0$ small
enough, and thus the conclusion is that either $p > p_{c}$, or
\eq{D} holds.


\newpage
\masect{ISC in Two Dimensions} \vspace{-.6cm} \label{sect:2D}
\masubsect{A Two-dimensional Construction}

\begin{figure}[ht]
    \begin{center}
    \leavevmode
    \epsfbox{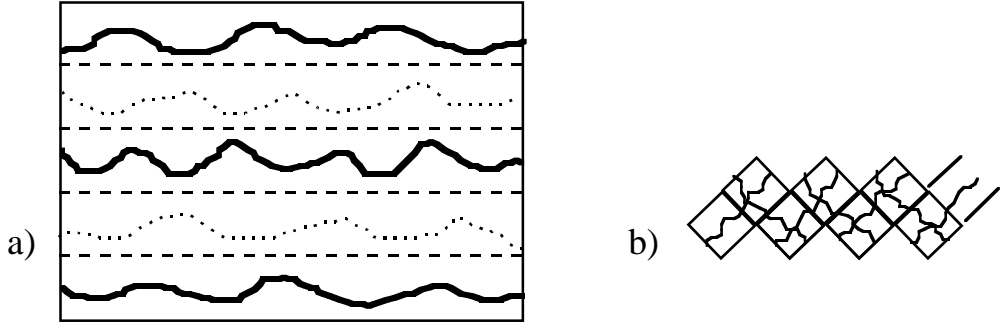}
  \caption{The elementary mechanism for multiple spanning clusters
  in 2D critical models: a) strips 
  traversed by connecting paths alternate with strips traversed by 
  separating dual paths (which avoid the realized connections), b) long
  connected paths are formed through the intersection of elementary 
  crossing events.}
\label{fig:2D}
\end{center}
\end{figure}

The situation in two dimensions is most amenable to
qualitative analysis.
The proof that there can be any number ($n$) of spanning clusters,
with probability which does not vanish as $L \to \infty$,
is elementary: partition the rectangular region
into  $(2n-1)$ parallel strips, and note that the event
occurs under the scenario depicted in Figure~\ref{fig:2D}.a.  Based on the 
product of probabilities of the $(2n-1)$ independent events, one gets:

\bea
K_{L}(n) &\equiv & \P\left(
   \begin{array}{l}
\mbox{the square $[0,L]^2$ is traversed (left $\leftrightarrow$ right)} 
\\ 
   	\mbox{by at least $n$ distinct [spanning] clusters}
   \end{array}
  \right) \  \ge \nonumber  \\   \nonumber  \\  
  & \ge & \  \P\left(
\begin{array}{l}
 \mbox{\small the strip $[0,L]\times 
            [0,L/(2n-1)]$ is traversed in the} \\
	\mbox{long direction by a spanning cluster [dual cluster]}
\end{array}  \right)^{2n-1}    \nonumber  \\   \nonumber  \\  
  &  \ge & \ [h_2(2n-1)]^{2n-1} \ > 0
\label{hoho}
\eea
where the concluding step is by  the
aforementioned theorem of Russo \cite{R} 
and Seymour and Welsh \cite{SW}
that  $h_2(t)>0$ for all $t$.

We view the above argument as part of the rich legacy of the
RSW theory.  This term refers  to a versatile
method for construction which in combination with
other insights has been employed
for a variety of far less elementary results:
non-percolation at $p_c$ \cite{R2}, sharpness
of the phase transition \cite{K1}, support for the
scaling theory \cite{K2} (all the above for 2D models),
some rigorous real space renormalization arguments \cite{ACCFR}
and some extension to higher dimensions \cite{ACCFR}.
(Extension to the continuum 2D random Voronoy cell percolation model 
is worked out in \cite{Alex}.)

Our main goal in this section is to present large--deviation
estimates for the probability that the number of clusters
spanning $[0,L]^{2}$ exceeds (or equals) a large number $n$.   

\masubsect{Bounds for the Number of ISC in 2D}

It is easy to be mislead about the large deviation asymptotics.
By an application of the van den Berg -- Kesten 
inequality 
\begin{equation}
	K_{L}(n) \ \le \ \left[ R_{L,1} \right]^n  
	 \ \le \ e^{- |\ln h_2(1)| \ 
	n }   \  ,
\label{sq.crossing}
\end{equation}
with $h_2(1)=1/2$ in the self--dual case.  Numerical 
results seem consistent with exponential decay in $n$
(of the form $e^{-Const.\ n}$) \cite{Sen}. 
Nevertheless the decay rate is different (faster).

\remark 
The inequality of van den Berg and Kesten \cite{vdBK}  states that for
independent percolation 
(or, generally, systems of independent variables)
the probability of the {\em disjoint occurrence} of two, 
or more, events 
(e.g., the existence of two separate connecting paths) is dominated
by the product of their probabilities.  The statement
is not obvious since {\em disjoint occurrence} does not 
refer to a-priori specified separate regions.  

The faster--than--exponential decay is caused by the 
mutual exclusion, which limits the clusters to reduced regions.
(For the lower bound seen in \eq{hoho} the clusters are produced in
disjoint strips).

\begin{thm}
In planar percolation models, at $p=p_{c}$ the 
probability that $[0,L]^2$ is traversed (``left $\leftrightarrow$ right'')
by at least $n$ separate clusters satisfies [for $n>>1$]
\begin{equation}
	K_{L}(n) \   \left\{
   \begin{array}{cc} \ge & A \ e^{-\alpha \ n^{2}} \\
 					 \le & \  \ e^{-\alpha' \ n^{2}}
    \end{array}   \right.   \  , 
\label{n}
\end{equation}
where the bounds involve different positive constants.
\end{thm}
\remark $Prob(\mbox{ exactly $n$ crossings })$  behaves just as
$Prob(\mbox{ at least $n$ crossings }) \equiv K_{L}(n)$, since the 
former is $K_{L}(n) - K_{L}(n+1)$, and (by the vdB-K inequality)
$K_{L}(n+1) \le K_{L}(1)\cdot K_{L}(n) \ (\  = K_L(n)/2 \  )$ .

\noindent{\bf Proof:}   The lower bound in \eq{n} is obtained through
the construction discussed above, supplementing \eq{hoho} with
the known fact that $h_2(t) \ge [\approx]  e^{-\alpha t}$ 
(proven by constructing a chain of 
$1\times 2$ brick  crossing events;  as  
seen in Figure~\ref{fig:2D}.b  \cite{R,SW,R2,ACCFR}).

The upper bound is slightly more involved.  Let us first ask
how many distinct spanning clusters will be seen in a tall
strip $L_{h} \times L_{v}$ ($L_{v} > L_{h}$).  
It is natural to expect that the number, call it $ N_{L_{v},L_{h}} $
will typically 
be of the order of $L_{v}/L_{h}$, 
since, roughly,  there is a finite number of 
distinct crossings in each of the squares comprising the tall stack.
This motivates the following bound, which is proven below using 
Lemma \ref{crossing.lm},
\begin{equation}
 \P\left( N_{L_{h},L_{v}}\ge u \frac{L_v}{L_h} \right) \ 
   \le  \ e^{ - f(u) \  L_v/L_h}
	\label{expN}
\end{equation}
with  $f(u)>0$ for $u$ large enough.

For a moment, let us assume \eq{expN}.  For our 
analysis, we fix $u$ at a value for which $f(u)>0$; the optimal 
choice being one maximizing $f(u)/u^2$ (one may note that 
$f(u)=0$ for $u<1/2$).

To estimate the multiple--spanning probability $K_{L}(n)$ 
(for $u < n < L$), 
we partition the $L\times L$ square into $n/u$ vertical strips,
so that the height/base ratio satisfies 
$ n = u L / L_{h}$.  If the square is spanned by 
at least $n$ disconnected clusters, then that is also the case
for each of the vertical strips.  (Though the converse need 
not be true.)  These being independent events, 
the probability is bounded by:
\bea
	K_{L}(n) \ & \le & \ 
	\left[  e^{-f(u)\ \frac{n}{u} }\right]^{n/u} \nonumber \\
	& \le &   e^{-  \frac{f(u)}{u^2}\ n^2 } \  ,
	\label{expNN}
\eea
which concludes the proof of \eq{n}.

To prove the large -- deviations estimate \eq{expN} we first 
derive the following related statement.

\begin{lem}  
\begin{equation}
	\E_{p_c}\left(e^{t \ N_{L_{h},L_{v}}} \right) \ \le \ 
	e^{\phi(t)\frac{L_v}{L_h} }
\end{equation}
with $\phi(t) < \infty$ at least for $t$ small enough.\\ 
{\em [A - posteriori, we'll find $\phi(t)$ to be finite for all $t$.]
$\E(--)$ represents expectation value.}
\label{crossing.lm}
\end{lem}
\proof 
To avoid more complicated analysis, we shall take advantage
of a convenient shortcut.  For each length $L$, let $\tilde{N}_L$ 
be the (random) number of distinct connected clusters 
connecting the left face of a square with the vertical 
line which includes the right face,
allowing the paths to meander into the entire
vertical strip  containing the square.   
We claim that there is a function $\phi(t)$, 
finite for $t$ small enough, such that:
 \bea
  \E_{p_c}( e^{t \ \tilde{N}_L } ) \ &=& \ 
 	         1 + t \int_{0}^{\infty} e^{tx}\ 
    Prob_{p_c}( \tilde{N}_L \ge  x)  \ dx    \nonumber \\
 	       & \le & 	e^{\phi(t)}   \  .
 \eea 

The claim is based on two observations: 
i. $Prob( \tilde{N}_L \ge k ) \le 
Prob( \tilde{N}_L \ge 1 )^{k}$ (using the vdB--K inequality),
and ii.  $Prob( \tilde{N}_L \ge 1)$ is  uniformly $< 1$
(a simple construction using the RSW theory).

A moment of reflection shows that the random variable $N_{L_h,L_v}$
is dominated, in the stochastic sense, by the sum of 
$L_{v}/L_{h}$ independent copies of $\tilde{N}_{L_h}$   [that is the 
short-cut.]   To see this, let us 
partition the strip's left vertical edge (of height $L_v$) into 
intervals of length $L_h$, and for each integer 
$1 \ \le k \le L_{v}/L_{h}$ let $N_k$ be the number of spanning 
clusters whose lowest intersection with the strip's left edge
falls within the $k$-th interval (counting from below).

The conditional distribution
of $N_k$, conditioned on \mbox{\{$N_n$ : $n < k$ \}}, can be 
studied by further conditioning on the exact location of the previously
counted clusters.  The net effect of the conditioning is an 
excluded volume.  Regardless of it, the conditional 
distribution of  $N_k$ is dominated by that of $\tilde{N}_{L_h}$, 
since the latter is recovered through the addition of two connected 
regions (one above the free region and one below) 
in which further clusters may be found. We rely here on the fact that
in 2D, the extra connections
cannot diminish the number of spanning clusters by joining different
ones, since they can reach only  the ``lowest'' 
(or the ``highest'') cluster.  
The most relevant implication of the stochastic--domination relation is: 
\bea
		\E_{p_c}\left(e^{t \ N_{L_{h},L_{v}}} \right) \ &\le& \ 
		\E_{p_c}\left(e^{t \ \tilde{N}_{L_h } }
		     \right)^{L_{v}/L_{h}} \nonumber \\
		& \le &  	e^{\phi(t) \frac{L_v}{L_h} }  \ ,
	\label{end.lm}
\eea 
as claimed in Lemma \ref{crossing.lm}.
(I thank H. Kesten for correcting the first draft's definition of 
$\tilde{N}_{L_h}$.)

The remaining path from \eq{end.lm} to \eq{expN} involves the
 Chebyshev estimate:
\begin{equation}
	Prob_{p_c}\left( N_{L_{h},L_{v}}\ge u \ 
          \frac{L_v}{L_h} \right) \ \le \  
 \E_{p_c}(e^{t \ N_{L_{h},L_{v}}}) \ e^{-t \  u \ L_{v}/L_{h}} \ 
	 \le \ e^{-[t u - \phi(t)] L_v / L_h}  \  ,
\end{equation}
which shows that \eq{expN} holds with $f(u)$ the Legendre transform:
\begin{equation}
	f(u)\  = \sup_{t}[ t u - \phi(t) ] \   \  . 
\end{equation} 
Since there are values of $t$ for which $\phi(t) < \infty$, $f(u)>0$ 
for $u$ large enough.

One may note that the results will not change in a qualitative 
way if instead of counting distinct clusters one
counts disjoint paths.
 
\begin{figure}[ht]
    \begin{center}
    \leavevmode
    \epsfbox{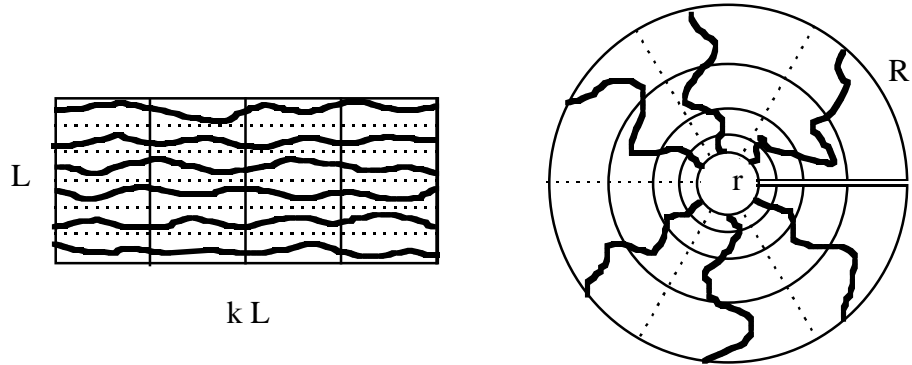}
  \caption{Two multiple--crossing events which are related by the 
  map $z \to r e^{2\pi z/L}$ (with $R / r \ = \ e^{2\pi k}$ and $L,
  \ r \ >> $
   {\em lattice spacing}).  Qualitatively, our bounds are 
  invariant under such conformal maps.}
\label{fig:conformal}
\end{center}
\end{figure}

Before we turn to the discussion of the spanning clusters in high 
dimensions, let us note that the arguments 
used to derive Theorem 3, provide also information on two other 
questions (depicted in Figure~\ref{fig:conformal}).  
The first relates to the {\em rate of exponential decay} 
for the probability that a {\em long} strip ($[0,kL] \times [0,L]$)
is traversed by $n \ (<L)$ distinct spanning clusters.  
We find that the rate of decay in the aspect ratio $k$
(the ``mass'' in field--theoretic jargon) is of the order of $n^2$:
\be
\P\left(
   \begin{array}{l}
\mbox{the strip $[0,kL]\times[0,L]$ is traversed (in } \\ 
   	\mbox{ its  long direction) by  $n$ distinct clusters}
   \end{array}
  \right) \ \ 
    \left\{
   \begin{array}{cc} \ge & A \ e^{-\alpha n^{2} \cdot k} \\ 
   \mbox{ }  \\  
 					 \le &  \ e^{-\alpha' n^{2} \cdot k}
    \end{array}   \right.   \  , 
\label{kstrips}
\ee
where the focus is on $k>>1$.  The proof is just the argument of 
Theorem 3, carried for rectangles rather than squares.

The second statement concerns the {\em power law} for the decay 
of the probability that an annulus is spanned by $n$ distinct clusters.
This time, it is the characteristic exponent which depends on $n$:

\be
\P\left(
   \begin{array}{l}
\mbox{the annulus $\{\ x\in R^2 \ : \ r < |X| < R \ \}$ } \\ 
   	\mbox{is traversed by $n$ distinct clusters}
   \end{array}
  \right) \ \ 
    \left\{
   \begin{array}{cc} \ge & A \ (r/R)^{-\alpha n^{2} /(2\pi)} \\
   \mbox{ } \\  
 					 \le &  \ \  (r/R)^{-\alpha' n^{2} /(2\pi)}
    \end{array}   \right.   \   
\label{circles}
\ee
where it understood that $1 \le n \le r/\mbox{\em lattice spacing}$.
As explained in Figure~\ref{fig:conformal}, the events seen in
\eq{kstrips} and \eq{circles} are related by a conformal map.
This map indicates also how to transcribe our proof of 
Theorem 3 into a proof of \eq{circles} for an annulus with a cut.
It is also easy to adapt the argument to the full annulus and,
correspondingly, to the rectangle with periodic boundary conditions. 

The constants $\alpha$ with which the above equations are proven are 
not identical, though it is natural to conjecture that the asymptotic 
values (in the sense seen below) are the same, and agree with the limit:
\begin{equation}
	\lim_{n \to \infty}  \lim_{L\to \infty}
	     \frac{1}{n^2}\log K_L(n) \  (= \ \alpha_{(asymp.)} \ ?) \ . 
\label{alpha}
\end{equation} 
It seems plausible that the exact value could be determined 
by methods based on the Coulomb gas representation \cite{Nie_rev}  
and / or conformal field theory  \cite{Car}.
Further discussion of the purported conformal invariance of the 
scaling limit can be found in references \cite{Car_rev,LPS,Aiz_IMA}. 

\bigskip

\bigskip
 
\masect{Above the Upper Critical Dimension} \vspace{-.6cm}
\masubsect{The Relevant Condition}

We shall now prove that in contrast with the situation in 2D,
above the upper critical dimension spanning clusters are sure to 
occur, and they proliferate 
in the way described in the introduction.  
A convenient sufficiency 
condition is that: \be d>6 \ee \noindent {\em and} the connectivity 
function defined by \eq{eta} (at $p=p_c$) satisfies: \be \tau(x,y) \  
\lg \ Const.  / \ |x-y|^{(d-2+\eta)}
  \label{t-c}
\ee
with
\be
 \eta = 0 \, .
 \label{eta0}
\ee

In this article, the relation ($\lg$) seen in \eq{t-c}
means that
\be
\tau(x,y) \ \ \left\{
   \begin{array}{cc} \le & C \ / \ |x-y|^{(d-2+\eta)} \\
 					 \ge & C'/ \ \ |x-y|^{(d-2+\eta)}
    \end{array}   \right.
\ee
with a pair of possibly distinct constants, $0<C' \le C< \infty$.

It is generally expected that $d>6 \Longrightarrow \eta=0$.
As mentioned in the introduction, there are mathematical results 
offering partial support to this claim, however some gaps remain 
\cite{HS}.
The statement which is established in ref. \cite{HS}
for high dimensional models is slightly weaker
than \eq{t-c} and \eq{eta0}, but sufficient for the ``triangle
condition'' of ref. \cite{AN} 
(``$\nabla$--diagram''($p_c$) $<\infty$).
A strategy which was successful in the derivation of some 
other results about the critical behavior in high dimensions 
was to base the analysis on such a weaker assumption
(proving ``$\nabla$-condition'' $\Longrightarrow$
$\gamma=1$ \cite{AN}, $\beta = 1$ and $\delta=2$ \cite{BA}).
We do not pursue that track here.  

\masubsect{The Boundary Conditions}

The number and the sizes of the connected clusters 
in a finite volume are affected by the boundary
conditions (b.c.). (I thank C. Borgs and J. Chayes for 
calling my attention to this point.)  The possibilities include:
\begin{itemize}  
\item {\em Free \/} b.c. ---  sites can be
connected only by paths within the specified volume, $\Lambda$. 
\item {\em Bulk \/ } b.c. ---  the given region $\Lambda$
is viewed as part of a much larger domain, taken to be all of $R^d$.  
Sites in $\Lambda$ can be 
connected by paths which stray outside of that region.
\item {\em Periodic \/ } b.c. --- in the well familiar sense.
\end{itemize}
(Omitted here are the {\em Wired\/} b.c., for which $N_{Wired}\equiv 1$.)

The obvious relations for the numbers of [left-right spanning] clusters, 
are:
\begin{equation}
	N_{Free} \ \ge \ N_{Bulk} \ , \ N_{Per} \ \ 
\end{equation}
with opposite inequalities for the maximal cluster size.

The terminology we adapt here is based on the {\em Bulk} b.c..
This choice facilitates the presentation of the main ideas, given
that our starting point is an explicit assumption concerning
the two point function in the bulk.
With additional information on the two point function, the method
should be applicable also to the other b.c..

\noindent {\bf Question:} 
In a sense made more precise below, we shall show that the 
maximal cluster size (Bulk b.c.) scales as $|C|_{max}\approx L^4$.  
Concerning the other boundary conditions:  we expect that 
the Free b.c. scaling is similar, in line with the prevailing theory
\cite{AGK, AGNW}, but we venture the guess that
the Periodic case is significantly different, possibly scaling 
as $|C|_{max}\approx L^{3d/2}$.  The latter is the scaling law
for the complete graph \cite{ER} (see also \cite{BCKS}).   
Our guess is motivated by an 
analogy with a phenomenon which has been noticed for the
$\phi^4_{d}$ field theory above its upper critical dimension \cite{BZ}.
Is the guess correct?  It will be of interest to see numerical work on 
this question.

\masubsect{The Number ($L^{d-6}$) and Size ($L^{4}$)
 of the  Spanning Clusters}

The geometric setup we use is that of the d-dimensional cube, denoting
$\Lambda \equiv \Lambda_{L} = [-L,L]^d$, and 
$\partial \Lambda_{- (+)} = \{x \in \Lambda_L \ | \ x_1 = -L (+L) \}$
(the ``left'' and ``right'' boundaries).
The symbol $\C$ would range over the connected clusters in $\Lambda$ 
(connected in the ``bulk b.c.'' sense).  

{\em Spanning clusters} are defined now as 
those intersecting both $\partial \Lambda_{-}$ and 
$\partial \Lambda_{+}$.  It can, however, be easily seen that none of 
the results will change (except for constants) if that notion is 
modified by declaring as spanning only those clusters which reach {\em each}
of the $2d$ boundary faces.   
 
The number of spanning clusters (which is a random variable) is denoted
by $N_L$.  We shall also be interested in the number $N_{L,W}$ 
of clusters spanning  
$\Lambda_L$ which are visible in a much smaller window $B_W= [-W,W]^d$, 
with $1$ (=lattice spacing) $  << W << L$:
\be
N_{L,W} = \mbox{card}
\left\{ \C \subset \Lambda_{L} \ | \ \C  \cap \partial 
\Lambda_{-} \ne \emptyset, \ 
 \C  \cap \partial 
\Lambda_{+} \ne \emptyset \mbox{ and } \C  \cap B_W \ne \emptyset 
\right\} \  .
\label{defNLW} 
\ee
(the location of the window within $\Lambda$ will not play an
important role.)

The main results in this section are the two Theorems found below.

\begin{thm}
In any dimension  $d>6$, if \eq{t-c} holds with $\eta=0$, then, as 
$L\to \infty$: \\ 
\noindent 1)  The spanning probability (at $p=p_c$) tends to $1$.\\ 
\noindent 2)  The probability of observing a spanning cluster within
the window $B_{W}=[-W,W]^d$ also tends to $1$, provided 
\be
\frac{W}{L^{2/(d-4)}} \ \Ltoo \infty   \ .
\label{W-L}
\ee

\noindent 3)  The numbers of spanning clusters (in $\Lambda$ and 
those observed in $B_W$) satisfy:
\be
\P\left( N_{L} \ \ge \ o(1)L^{d-6} \right) \ \Ltoo \ 1 \ .
\label{4.a}
\ee

\be
\P\left( N_{L,W} \ \ge \ o(1)\frac{W^{d-4}}{L^2} \right) \ 
\parbox[t]{.5in}
        {   $\longrightarrow$    \\ ${\scriptstyle L,\ W (\le L) \to \infty}$}
         \ \ \ 1 \ .
         \label{4.b}
\ee

\noindent in the sense that the statements are true for any function 
of L (and W), denoted here $o(1)$, which tends to $0$ as $L, W \to 
\infty$. \\ 
\label{thm4}
\end{thm}

The proliferation of the spanning clusters bears
some relation to their low (``fractal'')
dimension.
It is often stated that their dimension reaches $D=4$ 
and remains at that level, as the lattice dimension
is increased over the  upper critical value $d=6$
(\cite{AGK, AGNW, Con}).
The following statement offers some rigorous support to this claim.

\begin{thm}
In $d>6$  dimensions, assuming the power--low behavior (\eq{t-c}) 
with $\eta=0$,
\be
\P\left( \max_{\C \subset \Lambda_L}{}^{[(sp)]}\  |\C| \
\begin{array}{ll}
	\le &  \ c \ \log L\ \cdot L^4   \\
	\ge & \ o(1) \ L^{4}
\end{array}
  \right) \ \Ltoo 1
  \label{4.c}
\ee

and 
\be
\P\left( \max_{\C \subset \Lambda_L}{}^{[(sp)]}\  |\C| \cap \partial 
\Lambda_L \
\begin{array}{ll}
	\le &  \ c \ \log L\ \cdot L^3   \\
	\ge & \ o(1) \ L^{3}
\end{array}
  \right) \ \Ltoo 1
 \label{4.d} 
\ee
where  [(sp)] represents an optional restrictions to spanning clusters, 
and $o(1)$ is to be interpreted as in Theorem \ref{thm4}.
\label{thm5}
\end{thm}

In the derivation of these results we make use of the diagrammatic
bounds of ref. \cite{AN}.  
The most elementary of these is the tree--diagram
bound on the k-point connectivity function, which directly yields:

\begin{lem}  In any dimension, assuming the power-low behavior
\eq{t-c}, for any $k\ge 2$
\be
\E_{p_c}\left(|\C|_{max}^k \right) \ \le \
\E_{p_c}\left(\sum_{\C \subset \Lambda_{L}} |\C|^k \right) \ \le \ 
        \ k!  \ C_d^k \cdot 
L^{d-6+3\eta} \cdot L^{(4-2\eta)k}
\label{C-moments.1}
\ee

\be
\E_{p_c}\left(\max_{\C \subset \Lambda_L}|\C \cap \partial 
\Lambda_{-}|^k \right) \ \le
\E_{p_c}\left(\sum_{\C \subset \Lambda_L} |\C\cap \partial 
\Lambda_{-}|^k \right) \ \le \ 
        \ k!  \ C_d^k \cdot 
L^{d-6+3\eta} \cdot L^{(3-2\eta)k}
\label{C-moments.2}
\ee
and
\be
\E_{p_c}\left(\sum_{\C \subset \Lambda_{L}} 
|\C\cap B_W|\ |\C\cap \partial \Lambda_{-}|\ |\C\cap \partial 
\Lambda_{+}|  \right) \ \le \ 
        \ Const.\  W^d L^{4-3\eta}   \ \ .
\label{C-moments.3}
\ee
\label{lm2}
\end{lem}
 
Somewhat less immediate are the {\em lower bounds} seen 
in the following auxiliary statement.

\begin{lem}  In any dimension $d>6$, if \eq{t-c} holds with $\eta=0$, 
then for $k=1,\ 2$
\be
\P\left( \sum_{\C \subset \Lambda_{L}} 
 |\C\cap \partial \Lambda_{-}|\ 
 |\C\cap \partial \Lambda_{+}|  \ \
		\begin{array}{ll}
			\le & \ 1/o(1) \\
			\ge & \ o(1) 
		\end{array}
\times \   L^{d-6}\ L^3 \ L^{3\cdot k}  \right) \ \Ltoo \ 1 
\label{typical.1}
\ee
and
\be
\P\left( \sum_{\C \subset \Lambda_{L}} |\C\cap B_W|\ 
 |\C\cap \partial \Lambda_{-}|^k\ 
 |\C\cap \partial \Lambda_{+}|  \ \
		\begin{array}{ll}
			\le & \ 1/o(1) \\
			\ge & \ o(1) 
		\end{array}
\times \   W^d L^4   \right) \ \Ltoo \ 1 
\label{typical.2}
\ee
\label{lm3}
[where we read $W^d L^4$ as $\frac{W^{d-4}}{L^2} \ W^4\ L^3\  L^3$].
\end{lem}

Before deriving the two Lemmas (in the next subsection), let 
us go over the deductions of the 
Theorems (\ref{thm4} and \ref{thm5}), from Lemma~\ref{lm2} and 
Lemma~\ref{lm3} . 

\noindent{\bf Proof Theorem \ref{thm5}\/ {(\em assuming Lemma \ref{lm2}
and \ref{lm3})}:} \\
We deduce that 
$|\C|_{max} \equiv \max_{\C \subset \Lambda_L} |C\cap \Lambda_L| $ 
is typically not much larger than $L^{4}$, 
from the rate of growth of the moments described in \eq{C-moments.1} 
and  \eq{C-moments.2}, 
combined with the Chebyshev inequality (see \eq{chebyshev} below): 
\be \P\left( |\C_{max}| \ge 
\alpha(L) \cdot L^\lambda \right) \ \le \ \frac{\E\left(|\C_{max}|^k 
\right) }{\alpha(L)^k \cdot L^{k\lambda}} \ .   
\ee
A comparison with \eq{C-moments.1}  shows that a natural choice is 
$\lambda = (4-2\eta)$ and $\alpha(L) = a \log L$.   
Optimizing over $k$ 
(chosen so that $k \approx \alpha(L)/C_d$ ), one learns that 
\be
\P\left(|\C_{max}|\le a\ \log L\cdot L^{(4-2\eta)}
 \right) \Ltoo 1 \ ,
\ee
provided $a > C_d (d-6+3\eta)$.
This proves the upper bound on $|\C|_{max}$ claimed in \ref{4.c}, in 
a somewhat more general form.
A similar use of the Chebyshev inequality leads from 
\eq{C-moments.2} to the conclusion that typically $max_{\C \subset 
\Lambda_L}|\C\cap \partial \Lambda_L| \le a' \log L L^{3-2\eta}$, as 
claimed in \eq{4.d}.

To deduce that  $|\C|_{max}$ is typically not much smaller than the 
above upper bound ($L^4$), we use :
\be 
|\C|_{max} \ \ge \ |\C^{(sp)}|_{max} \ \ge \ \frac{\sum |\C\cap \Lambda |\ 
 |\C\cap \partial \Lambda_{-}|\ 
 |\C\cap \partial \Lambda_{+}|}
 {\sum |\C\cap \partial \Lambda_{-}|\ 
 |\C\cap \partial \Lambda_{+}|}
\ee
(where the sums are over the connected clusters of $\Lambda = \Lambda_L$)
combined with the information about the typical values, provided by  
\eq{typical.1} with $k=1$,  and \eq{typical.2} at $W=L$.
Likewise, the lower bound claimed in \eq{4.d} on 
$\max_{\C \subset \Lambda_L}{}^{[(sp)]}\  |\C| \cap \partial 
\Lambda_L$ follows by comparing the lower bound in \eq{typical.1}
for $k=2$ with the upper bound in the same equation for $k=1$. 

\noindent{\em Proof Theorem \ref{thm4}\/ {\em (assuming Lemma \ref{lm2}
and \ref{lm3})}:}  
To show that typically there are at least $o(1) L^{d-6}$  spanning clusters 
we
use:
\be
N_L \ \ge  \ 
\frac{\sum |\C\cap \partial \Lambda_- | \ |\C\cap \partial \Lambda_+ |}
{\max |\C\cap \partial \Lambda_-| \ |\C\cap \partial \Lambda_+|}  \ .
\ee
Typically, the numerator is $\ \ge o(1)\ L^d$ (by Lemma \ref{lm3})
while the denominator is \\
 \mbox{$\le 1/o(1)\ L^{2\cdot 3}$} (by 
Theorem~\ref{thm5}).  Hence 
$N_L \ \ge \ o(1)\ L^{d-6}$, as claimed in \eq{4.a}.

At the risk of sounding too repetitive, for the number of 
the spanning clusters seen in the window  $B_W$ we use:
\be
 N_{L,W}  \ \ge \frac{\sum |\C\cap B_W| \ |\C\cap \partial \Lambda_- | \ 
  |\C\cap \partial \Lambda_+ |}
  {\max |\C\cap B_W| \max |\C\cap \partial \Lambda_- | \ 
  |\C\cap \partial \Lambda_+ |}  \ ,  
\ee
which leads to the conclusion that typically:
\be
 N_{L,W} \ge  o(1) \frac{W^d L^4}{W^4 L^3 L^3} \ = \ o(1) \frac{ 
 W^{d-4}}{ L^2} \  .
\ee
Under any conditions which guarantee $ N_{L,W} > 1$, we are assured 
that a spanning cluster intersects the box $B_L$.
This concludes the proof of Theorem~\ref{thm4}.  

A particular conclusion which seems worth emphasizing is:
\be
\P \left\{
\begin{array}{l}
	\mbox{the window $[-W,W]^d$ is connected to the boundary} \\
	\mbox{of the (much larger) box $[-W^\alpha,W^\alpha]^d$}
\end{array}
 \right\}  \Ltoo  1
\label{4.continuum}
\ee 

\noindent provided  $\alpha < (d-4)/2$ (assuming all along
that $d>6$, and $\eta=0$).

\masubsect{Derivation of the Moment Bounds}

We start from the elementary but important identity
\bea
\E\left( \sum \ |\C|^k \right)  \ &=
 &\ \E\left( \sum_{\C}\sum_{x_{1},\ldots,x_{k}} 
   \I[x_{1}\in \C]\ \ldots \I[x_{k}\in \C]  \right) 
 \nonumber \\
& = & \ \E\left( \sum_{x_{1},\ldots,x_{k}}
\I[\mbox{ $x_{1},\ldots, x_{k}$ belong to a common 
[spanning] cluster}]  
\right)  \nonumber \\
& = & \sum_{x_{1},\ldots,x_{k}}\tau(x_{1},\ldots, x_{k})
\label{c^identity} 
\eea
where $\I$[--] is an indicator function and 
$\tau_k(x_1,...,x_k)$ is the 
probability that the $k$ points are all
connected.

%
\begin{figure}[ht]
    \begin{center}
    \leavevmode
    \epsfbox{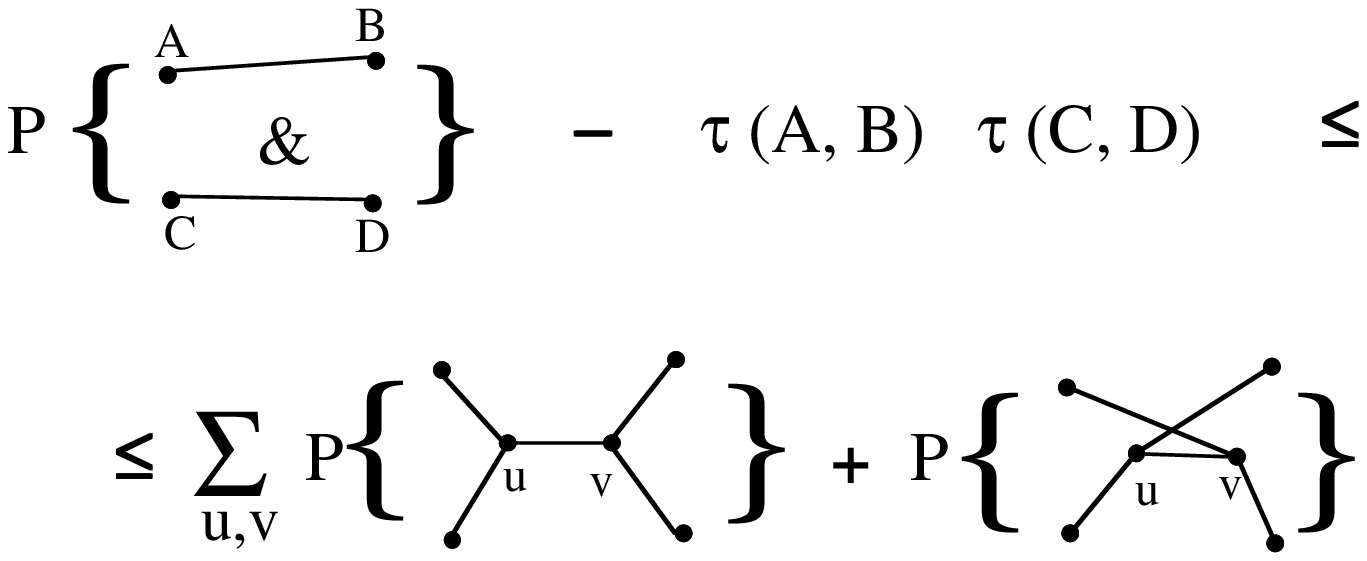}
\caption{
The inequality used for estimating the variance of $\sum |\C|^{2}$.
The truncation adds in the diagram two vertices and three lines.  
The result is an extra factor of $Const.  \ 
L^{2d}/L^{3(d-2+\eta)}=Const.\ /L^{d-6+3\eta}$, which above the upper 
critical dimension tends to $0$ for large separations.}
\end{center}
\label{fig:trunc}
\end{figure}
%

Next, the argument will  employ diagrammatic bounds on 
the  connectivity functions, following the 
approach presented in ref. \cite{AN}.
The technique has been further simplified through the development of 
the van den Berg -- Kesten 
inequality \cite{vdBK} mentioned in Section 3.
There are two principles whose
repeated application leads to Lemma~\ref{lm2} and Lemma~\ref{lm3}.
(We shall attempt to avoid obvious repetition of similar arguments.)
They are: \\
1) the {\em tree--diagram bound\/} \cite{AN}, and \\
2) the {truncation lemma\/} (Lemma~\ref{truncation-lemma}), proven below.

The tree--diagram bound \cite{AN} states 
that $\tau_{k}(x_{1},\ldots,x_{k})$ is
dominated by the sum of products of 
the two point function $\tau(u,v)$, arranged in the form of tree 
diagrams with the external vertices $ x_1, ..., x_k $. 
Its intuitive explanation is that in order for the given 
$k$ sites to be connected there has to be a connecting tree.  
The vdB-K inequality \cite{vdBK} 
(or the alternative argument which was originally used in 
ref.\cite{AN}) permits to 
bound the probability by the sum over the 
corresponding tree diagrams.
(The bound is not totally obvious since it involves sum over tree 
diagrams, and not tree graphs, i.e., 
all trees embedded in the lattice.
The latter sum is much too large.)

For $k=3$ the inequality is:
\be  
\tau_3(x_1, x_2, x_3) \ \le \
   \sum_u \tau(x_1,u) \  \tau(x_2,u) \  \tau(x_3,u)  \ .
\label{treegraph}
\ee
For general $k\ge 3$ there are $(2k-5)!! \le 2^{k} k!$ 
tree diagrams for 
each set of the external
vertices, and each diagram has  $(2k-2)$ 
vertices (external $+$ internal) 
and $(2k-3)$ lines.  For example, for the quantity seen in 
\eq{c^identity} the sum yields:
\be
\E_{p_c}\left( \sum \ |\C|^k \right)  \ \le \ k! \ C_d^k \cdot 
 L^{d(2k-2)} / L^{(4-2\eta)k} \ = \ k! \ C_d^k \cdot L^{d-6+3\eta} 
  \cdot L^{(4-2\eta)k}  \ .
\label{sumk}
\ee

A different argument is needed for the {\em lower bounds} seen in 
Lemma \ref{lm3}.  These rely on a more delicate estimate of the
difference between two comparable terms, which under the right 
conditions cancel up to a small remainder.  Following is that
auxiliary result.

\begin{lem}  For independent percolation on any graph,
\bea
0 &\le& 	Prob\left(
	\begin{array}{l}
	 \mbox{{\small $\{x_1,\ldots, x_n\} $ are all connected, and} } \\ 
		\mbox{{\small $\{y_1,\ldots,y_n\} $ are all connected}}
	\end{array} 	\right)  -  \tau_n(x_{1},\ldots, x_{n}) 
	\times \tau_n(y_{1},\ldots, y_{n})  \ 
	\le  \nonumber	\\  	 	
	&\le & 	Prob\left( \mbox{ $x_1,\ldots, x_n$, and $y_1,\ldots,y_n$
	belong to a common cluster} \right)  \ ,  \label{ntrunc}  
\eea
and, for $n=2$:
\bea
0\ &\le& \	Prob\left(
	\begin{array}{l}
	 \mbox{{\small $\{x_{1},x_{2}\}$ and $\{y_{1},y_{2}\}$}}  \\
		\mbox{{\small are pairwise connected}}
	\end{array} 	\right) \ - \ \tau(x_{1},x_{2}) 
	\times \tau(y_{1},y_{2})  
	\le  	\label{2trunc} \\  	 
	&\le & \sum_{u,v} \tau(x_{1},u) \ \tau(y_{1},u) \ 
	 \tau(v,u) \ \tau(v,x_{2})
	\ \tau(v,y_{2}) \ + 
	\ [y_{1} \leftrightarrow y_{2} \mbox{ permutation }] \ . \nonumber 
\eea
\label{truncation-lemma}
\end{lem}
{\bf Proof:} The positivity of the difference seen in \eq{ntrunc} is 
obtained by the standard monotonicity argument 
(the Harris -- FKG inequality 
\cite{Har, FKG}): the probability of the simultaneous connection of 
the two pairs is greater that the product of probabilities.  

In the  opposite direction, if the two collections of sites are
interconnected, then they are either in disjoint clusters or in a 
common cluster.  The probability of the first alternative is smaller 
than the product which is subtracted (by the vdB-K inequality 
\cite{vdBK}).  
Thus, we are led to \eq{ntrunc}.  

We could in fact make a stronger statement: by separating not just
the case that the two sets are in disjoint clusters, but the case
that they are disjointly connected, we are left with a smaller 
remainder.  In each of the configurations contributing to it
there is a tree subgraph connecting all the $2k$ vertices.  The tree is 
bound to have a link (a pair of internal vertices $\{u,v\}$) 
whose removal will split it into two subgraphs of equal numbers of 
vertices, and there is a further constraint that this 
link cannot separate the $x-$sites from the $y$-sites.  For $n=2$, 
that leads to \eq{2trunc}.  
 
\noindent {\bf Proof of Lemma \ref{lm2}} (following ref. \cite{AN}):\\ 
The first claim, \eq{C-moments.1}, is proven in
\eq{sumk}.  
The other statements of Lemma \ref{lm2} are derived similarly; 
e.g., for \eq{C-moments.3} one may start with:
\be
\E_{p_c}\left(\sum_{\C \subset \Lambda} 
|\C\cap B_W|\ |\C\cap \partial \Lambda_{-}|\ |\C\cap \partial 
\Lambda_{+}|  \right) \ = 
\ \sum_{x\in B_W,\ y\in \partial \Lambda_-,\ z\in  \partial \Lambda_-}
\tau_3 (x, y, z)  \  ,
\ee
and follow by an application of the tree--diagram bound, \eq{treegraph}.

%
\begin{figure}[ht]
    \begin{center}
    \leavevmode
    \epsfbox{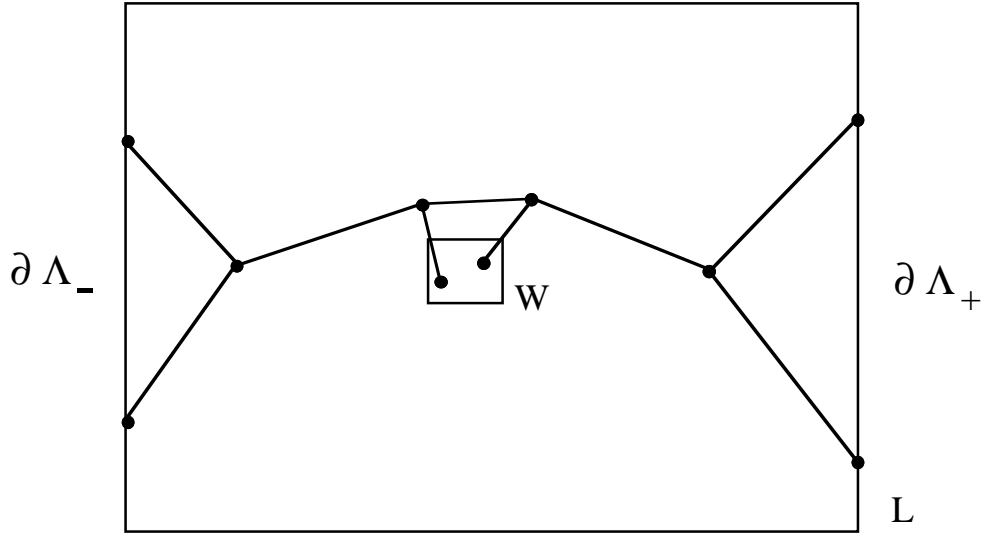}
\caption{The dominant tree--diagram in the bound on 
Var($\tilde{K}$).  As in Fig. 3, the truncation adds to the graph 
two extra sites and three extra lines, resulting here in the
multiplicative factor  $Const./ W^{d-6+3\eta}$. }
\end{center}
\label{fig:highd}
\end{figure}
%

\noindent {\bf Proof of Lemma \ref{lm3}:} \\
Let us consider first \eq{typical.1}.  The upper bound seen there
is a direct consequence of the value of the mean
and the Chebyshev inequality: 
\be Prob( X \ge t ) \  \le \  \frac{\E(X)}{t}  \  . 
\label{chebyshev}
\ee
A somewhat more delicate argument is needed to deduce that 
typically the sum is not much below the mean--value level.

As in $|\C|^2$ version of
the identity \eq{c^identity}, it is easy to see that:
\be
 \sum_{\C \subset \Lambda_{L}} 
 |\C\cap \partial \Lambda_{-}|\ 
 |\C\cap \partial \Lambda_{+}| \ = \   
 \sum_{x\in \partial \Lambda_-, \ y \in \partial \Lambda_+}
 \I[\mbox{ $x$ and $y$ are connected }] \
:= \ K \ ,
\label{K}
\ee
where $K$ is defined by the sum.
The mean value of $K$ is easily determined:
\bea
\E_{p_c}( K ) \ &=& \ \sum_{x\in \partial \Lambda_-,
 \ y \in \partial \Lambda_+}
\tau(x,y) \nonumber \\
& \lg & \mbox{Const.} \ L^{2 (d-1)} / L^{d-2+\eta} =
\mbox{Const.} \ L^{d-\eta}.
\label{4.15}
\eea

The corresponding expressions for the 
second moment, and for the variance,
are:
\be
\E( K^{2} ) \ = \ \sum_{ 
\begin{array}{l}
	x_1, x_2 \in \Lambda_-  \\
	y_1, y_2 \in \Lambda_+
\end{array}     }
\ Prob\left( \mbox{$x_{1}$ is connected to $y_{1}$, and $x_{2}$ is
         connected to  $y_{2}$ } \right)
\ee
and 
\bea
	Var(K) &=& \ \E( K^{2} ) - \E(K)^{2} \ = \  \   \\
	& = & \sum_{ 
\begin{array}{l}
	x_1, x_2 \in \Lambda_-  \\
	y_1, y_2 \in \Lambda_+
\end{array}     }   \left[
  Prob\left( 
\begin{array}{l}
	\mbox{$\{x_{1}, y_{1}\}$ and  $\{x_{2}, y_{2}\}$ } \\
	\mbox{are pairwise connected}
\end{array} \right)  - \ \tau(x_{1}, y_{1}) \times \tau(x_{2}, 
y_{2}) \right] \ .
	\nonumber 
\eea

Using \eq{2trunc} of the {\em truncation Lemma\/}, Lemma 
\ref{truncation-lemma},
we obtain the following estimate (where $\E$ should be read as
$\E_{p_c}$):
\bea
\E\left|\frac{K}{\E(K)} - 1\right|^2 \ &=& \
\frac{\E(K^2) - \left| \E(K) \right|^2}{\left| \E(K) \right|^2} 
\nonumber   \\
 &\le& \mbox{Const.} \ L^{2d} / L^{3(d-2+\eta)} \ = \
 \frac{\mbox{Const.}}{ L^{d-6+3\eta} } \ .
 \label{4.k}
\eea
In essence, the ``truncation'' results diagrammatically in 
the addition of {\em two sites} and {\em three lines}, which 
translates to the multiplicative factor of order $L^{2d} / 
L^{3(d-2+\eta)} =\ $ $\ L^{d-6+3\eta}$.
Under the right conditions, 
\eq{4.k} implies that only very seldom 
will $K$ differ  by a significant factor from the  mean $\E(K)$. 

We can proceed only under the assumption that $d-6+3\eta > 0$, for
otherwise the last bound still allows the typical values of $K/\E(K)$
to be arbitrarily small.  That however is ruled out
if the right side of \eq{4.k} is $o(1)$, in which case 
there is a constant $b>0$ (determined through \eq{4.15}) for which
\be
Prob_{p_c}\left( K \ \ge \
b\ L^{(d+2-\eta)} \right) \ \Ltoo \ 1  \ .
\ee
Since $\sum \mbox{}^{(sp)} \ |\C|^2 \ge K$ 
that implies the lower bound claimed in \eq{typical.1},
and thus concludes its proof.  

The derivation of \eq{typical.1} is very similar, with $K$ replaced by
$\tilde{K}$ defined by the following modification of \eq{K}
\be
 \sum_{\C \subset \Lambda_L} 
 |\C\cap B_W|\ |\C\cap \partial \Lambda_-|\ 
 |\C\cap \partial \Lambda_+| \ = \   
 \sum_{
 \begin{array}{c}
 	 x\in B_W   \\
 	y\in \partial \Lambda_- \\ 
    z \in \partial  \Lambda_+
 \end{array}          }  
 \I[\mbox{ $x,\ y$, and $z$ are connected }] \
:= \ \tilde{K} \ .
\label{K-tilde}
\ee
In the study of the variance $Var(\tilde{K})$, 
the role of equation \eq{2trunc} is taken by \eq{ntrunc} combined
with the tree diagram---bound.  
As before the ``truncation'' results diagrammatically in 
the addition of {\em two sites} and {\em three lines} (see 
Figure~\ref{fig:highd}),
 but now the
main contribution comes from diagrams with two internal sites within
distance of order $W$ from the window $B_W$.  That
translates to the multiplicative factor $Const.  \ W^{2d} / 
W^{3(d-2+\eta)} =Const.  / \ W^{d-6+3\eta}$.

\noindent {\bf Remarks:} 
{\em 1. } Purists may note that the above proof, that 
spanning is certain, actually requires 
only the weaker condition:
\be
d-6+3\eta > 0  \; ,
\label{d-6+3eta} 
\ee
under which the conclusion is that the number of spanning clusters
is typically greater than $o(1) L^{d-6+3\eta}$.  This can be turned 
around to say that in dimensions in which the spanning probability 
does not tend to $1$: $ \eta \le \frac{6-d}{3} $.  The importance of 
this improvement is dimmed by the fact that according to numerical 
estimates $\eta < 0$ for $d = 3, 4, 5$.

{\em 2. } It is natural at this point to enquire about an upper bound 
on the number of spanning clusters.  Comparing \eq{C-moments.1} with 
Theorem \ref{thm5}, we see that for counting clusters whose volume is 
comparable to the maximal the lower bound we got is sharp up to 
$\times L^{o(1)}$.  However our estimates will not detect the presence 
of a large number (larger that $L^{d-6}$, but not too large) of 
sufficiently thin spanning clusters.

\bigskip

\bigskip

\masect{The IIC, the ISC, and the Scaling Limit} \label{IIC-ISC}

\masubsect{The Microscopic and the Macroscopic Perspectives}

In this section we place the above results in the 
context of the interplay between the different scales 
on which the system can be viewed.

While percolation models are often presented on an infinite 
lattice, many of the interesting questions 
relate to finite, though large,
systems.  There are therefore at least two scales of interest 
(we focus on two but it will also be interesting to bring in a third, 
intermediate scale, which could allow a better mathematical expression 
of the renormalization group approach):
\begin{itemize}
\item[i.] The {\em microscopic scale,\/ } for which the unit length 
($a$)
is the lattice spacing, or the size of the dots in the random dot 
model.  This is the scale on which the elementary connections are 
defined.

\item[ii.] The {\em macroscopic scale,\/ } for which the unit length 
could be the system's width ($L$).  This is the scale of the clusters 
on which we focused.  Of particular  
interest are the clusters which connect two different boundary faces.
We generically refer to those as spanning clusters (in different 
geometries, see Figure~\ref{fig:cube}).
\end{itemize}

The situation in which $L >> a $ appears rather differently from the 
two perspectives.

On the microscopic scale the entire system appears vast, and it is 
mathematically advantageous to consider it infinite, and homogeneous 
(in the suitable sense).  The relevant limit is
$L \to \infty $.  The local structure of the very large clusters
can be studied on that scale but only if one first centers the viewing 
field on one of their sites.  That requires some effort, 
since at $p=p_c$ the density of the {\em union of macroscopic--scale
clusters} vanishes.  Thus is born the Incipient Infinite 
Cluster, which can be defined quite precisely through a number of 
procedures, most of them presumably yielding the same limiting object
(see below).

On the macroscopic scale, the situation $L/a >> 1 $ is expressed through
the {\em scaling limit}  $a \to 0$.
This limit lends itself naturally to the discussion of the 
higher symmetry, including possibly conformal invariance,
which seems to emerge on the large scales in the 
critical regime.  

When we count the number of the spanning clusters,
or other clusters of linear extent comparable with the system's size
(say diameter exceeding $0.8 L$, where 0.8 can be replaced by any 
fixed $0<s<1$),
we are obviously taking the macroscopic perspective.  

One should avoid being too dogmatic, as some of the 
features of the large clusters appear similar when one looks
either down from the macroscopic scale, or up from
the microscopic scale  (e.g., the structure of voids, discussed 
in the recent study  ref.~\cite{HASM}).
However, ignoring the above scale distinctions one runs the peril 
of pitfalls and paradoxes.   

An illuminating example is the following delightful paradox due to 
D. Stauffer \cite{St-paradox}.
There is an early rigorous result \cite{NS}
(even preceding the proofs of the uniqueness of the infinite cluster 
\cite{AKN,BK}) that in an infinite translation--invariant system 
(meeting a certain regularity condition)
the number of infinite clusters is almost surely constant:  
either $0, 1$ or 
$\infty$.  In a finite volume, the term ``infinite clusters''
ought to cover the spanning clusters.  Since infinity is ruled out,
and we know that at $p=p_c$ spanning clusters do occur, 
we are led to the conclusion that 
typical configurations of critical models, 
exhibit a unique spanning cluster.  
That however is in direct contradiction with 
Theorem~\ref{non-uniqueness}.  

The answer to Stauffer's paradox lies in a careful analysis of the 
relevant meaning of infinity.  Positive probability of there 
being an infinite cluster (on the short scale) means that for some 
fixed $K$ (large enough) the probability for observing
a cluster of size $R$ within the distance $K$ from the 
lattice origin (or alternatively from an ``averaged site'' 
in the large box)  
{\em does not tend to zero} in the limit $R, L \to \infty$.  
(For independent percolation one may take $R=L$; the 
distinction is needed for correlated percolation models where 
``thermalization'' may require 
the limit $L\to \infty$ to precede $R\to \infty$.)  
We leave it as an exercise to the reader to determine the 
finite-volume translation of the statement
that there are {\em more than two} infinite clusters in the infinite
volume limit.  

At $p=p_c$ no infinite 
clusters are seen in the infinite translation--invariant limit,
assuming that the percolation probability vanishes (as is
proven for 2D and for sufficiently high dimensions).  
Spanning clusters and other macroscopic--size clusters occur, but 
the density of their {\em union} tends to zero, and thus they leave 
no trace in the usual infinite volume limit.  (Their microscopic scale
separation also increases, and the IIC construction  
yields a unique infinite cluster.)
As we see, the paradoxical contradiction might be first ``corrected'' 
as not between one [infinite cluster] and a number greater than that
[of spanning clusters], but between zero and 
a number greater than $1$.  The resolution is through a finer 
appreciation of the difference in the scales on which the objects 
are observed, and defined.

That leaves us with another paradoxical observation:
{\em the infinite lattice represents just a small part of the 
finite system}.
This fact has been noticed before  \cite{SML}, when it was appreciated
that for ferromagnetic spin systems with the order parameter 
$M = \lim_{x\to \infty} |<\sigma_0 \sigma_x >|^{1/2}$ 
a distinction ought to be made
between  the Short -- Long -- Range Order parameter, defined
on the infinite system, 
and the Long --  Long -- Range Order parameter
for which the distance between the spins scales along with
the box size.
(Ipso facto, for the case considered in \cite{SML} the two order
parameters can nowadays be proven equal, but the distinction 
is of relevance for expectation values of quantities which are 
somewhat less local than the spins, as the Stauffer paradox 
demonstrates.)

In mathematical terms, a careful inspection of the notion of
measurability in product spaces also shows that the standard 
infinite system (e.g., the lattice) is in essence just the 
collection of all regions which remain at fixed microscopic--scale 
distances from the point represented by the lattice origin.  
Thus, again: the infinite system represents just a small part of the 
finite region studied in simulations.

Let us add that it is an interesting question of what mathematical
framework can be used to contain the macroscopic information of 
relevance in the limit $a\to 0$.
The spanning clusters and other macroscopic--scale clusters have 
the appearance of fractal objects.  However, it turns out that
the standard description of fractals does not capture the relevant 
information on the realized connections.  Thus a new formalism is 
required.  A proposal for the limiting description of the percolation
``Web'' is presented in \cite{A_Web} and work in 
progress.  It is  also of interest to take the double limit:  
$a \to 0$ and $L \to \infty$, which is needed if one wants
to ask about percolation in the scaling limit, on which more is said 
below.

\masubsect{Conjectured Properties of the IIC and the ISC}

Of the different constructions which were proposed for what one
may call the Incipient Infinite Cluster, our notion is closest to
that of Kesten \cite{K3} 
(another proposal \cite{CCD} is mentioned below).
The goal is to provide the mathematical description of the large 
clusters, down at the microscopic scale.  

One may guess that the following three
algorithms for the construction of a probability measure of 
the percolation 
model in which a chosen site is constrained to be connected to 
infinity, will have common limits.  The infinite cluster
seen there would be called the IIC, 
and the limiting measure will provide its 
probabilistic description.
\begin{itemize}
\item[a.] Generate the probability distribution for a random 
environment by a two step procedure. First pick a typical random 
configuration, and then center it relative to one of the sites on the 
large clusters, sampling with equal weights over all the sites 
connected to the boundary of $[0,L]^d$, possibly with a corrective
exclusion of the sites closer to the boundary than some $g(L)$, with 
$g(L)\to \infty$ for $L\to \infty$ (e.g., $g(L)= 0.1 L$). 
\item[b.] Take 
the conditional distribution conditioned on the origin being connected 
distance $L$ away, and let $L \to \infty$.  
\item[c.] Raise $p$ over 
$p_{c}$, condition on the origin being in the infinite cluster (which 
now exists), and let $p \searrow p_{c}$.
\end{itemize}  

For 2D, Kesten \cite{K3} proved the convergence and equality of 
the limits {\em b.\/} and {\em c\/}. 
Presumably, the method should cover also  {\em a.\/}   
(which seems to be the most efficient for simulations).
While the result 
was not yet extended to higher dimensions, it is expected to be true 
for general $d>1$.  Regardless of that, under the assumption 
$P_{\infty}(p_{c})=0$ one can prove that in each of the constructions, 
the limiting measure will exhibit exactly one infinite cluster (for 
$p=p_c$).  This uniqueness, however, is not the consequence of any of 
the uniqueness results cited above, since the measure(s) are neither 
regular nor translation invariant.

Another construction of IIC was suggested in 
ref.~\cite{CCD}, where an infinite cluster is formed by enhancing 
the the bond densities at a rate which tapers off as $\|x\| \to 
\infty$).  Presumably, the different constructions will agree 
in the further limit $a \to 0$ (the continuum limit).

One may be puzzled by the discrepancy between the multiplicity
of the Incipient Spanning Clusters 
(non-unique by Theorem~\ref{non-uniqueness})
and the uniqueness claimed for the IIC.  
The long answer was provided above. The short answer is
that the different ISC are at increasing distances apart (on the 
lattice scale).

We have also a natural conjecture concerning the 
scaling limits of the Spanning Clusters (not just at $p_c$).
We guess that for any $0<p<1$ one would find
just one of the three possibilities, which would be equally valid,
with probability which tends to one, 
for any bulk shape of the type seen in Figure~\ref{fig:cube}.
\begin{itemize}
\item[i.] No clusters attain size visible in the scaling limit.
\item[ii.] There is a unique macroscopic connected cluster 
filling the 
region densely (in the macroscopic--scale sense).  
\item[iii.] Multiple 
clusters with macroscopic--scale diameters will be seen in any 
rectangular region (regardless of its aspect ratio).
\end{itemize}

Existing results permit one to easily deduce:
\begin{eqnarray}
	p < p_c & \Longrightarrow & \mbox{alternative i.}
	\nonumber  \\  
    p > p_c & \Longrightarrow & \mbox{alternative ii.  
    (see Appendix C)}  \\  
    \mbox{for 2D:} \  p = p_c & \Longrightarrow & 
    \mbox{alternative iii.  (see Section 3)}
	\nonumber  
\end{eqnarray}
and in Section 2 it was established that
\begin{eqnarray}
	 \mbox{any $d >1$:} \ \ \  p = p_c & \Longrightarrow & 
	 \mbox{a restricted alternative iii.  } \ \ ,
\end{eqnarray} 
where the restriction is to rectangular regions with a 
conveniently small aspect ratio.  
Since the discussion refers to the scaling
limit, it seems most natural to expect the restriction to be
irrelevant in any dimension, though that is not proven here.
Such a result will amount to a significant extension
of the theory of Russo \cite{R} and Seymour and Welsh \cite{SW}, 
which is based on intrinsically 2D arguments.

\masubsect{Distinction between Type I and Type II Models}

We shall not discuss here the 
interesting question of what mathematical 
object will provide a suitable 
description of the structure emerging in the continuum limit.  
Let us, however, point out that a distinction ought 
to be made between two classes of critical models.   
We characterize them as follows.

\begin{itemize}
\item {\em Type I models:} The function
\begin{equation}
	\limsup_{L\to \infty} \P\left(
	\begin{array}{l}
		\mbox{the set $[-sL,sL]^d$ is connected}  \\
		\mbox{to the boundary of $[-L,L]^d$}
	\end{array}
	   \right) \ =  \tilde{h}(s) \  
\label{5.c}
\end{equation}
is strictly less than one, for some $0<s<1$; in which case 
$\lim_{s\to \infty} 
\tilde{h}(s)=0 $.

\item {\em Type II models}: 
\begin{equation}
	 \P\left(
	\begin{array}{l}
		\mbox{the set $[-sL,sL]^d$ is connected}  \\
		\mbox{to the boundary of $[-L,L]^d$}
	\end{array}
\right) \Ltoo 1 \ ,
\label{5.d}
\end{equation}
for any  $0 < s < 1$.
\end{itemize}
\noindent ({\em Question:\/} is all else ruled out?  
[Presumably --- yes.])

When viewed on the macroscopic scale, models of Type I exhibit
many clusters of still visible size, but none of 
them is  infinite.   The scaling limit of ISC in such models can be 
formulated along the lines discussed in ref. \cite{A_Web} 
(and work in progress).

In Type II models, the clusters visible on the macroscopic scale are 
qualitatively different than in Type I.  
Translated to the continuum scale, condition (\ref{5.d}) can be read as
saying that for any $R, \ \epsilon >0 $ 
the probability that an $\epsilon$ neighborhood of a given site is
connected distance $R$ away tends to $1$, in the continuum limit.

The high--dimensional percolation models 
discussed in Section 4 are of Type II.
One may gain a great deal of insight about their continuum limit from 
\eq{4.continuum}, and \eq{4.b}.   Consider the situation in which
the lattice spacing is $a \to 0$, and we observe the connected clusters
which intersect a cubic region of fixed continuum size, $r$.  On the 
lattice scale, the size of the box is 
\be
W \ = \ r/a  \ \  (\to \infty ) \  \ .
\ee
Equation (\ref{4.continuum}) implies that for any $0< \delta (< d-6)$
with predominant probability we shall see a cluster reaching distance 
$ R \ = \ a \ W^{(d-4-\delta)/2} \ = a^{-(d-6-\delta)/2}\ 
r^{(d-4-\delta)/2}$ ($\to \infty)$ away from the $r$-cube 
($r$ and $R$ are expressed 
in continuum  units).   In fact, the number of such clusters diverges,
according to \eq{4.b}.  If the bound produced there on 
$N_{L,W}$ is a correct indication of the actual value, then not all 
the cluster which span the $R$-box and intersect the $r$-box will 
intersect neighboring cubes of size $r$ (the fraction which does, scales 
as $r^{d-4}$).  The picture which 
is suggested by the (partial) results presented here is that there is 
a growing number of clusters reaching increasing distances, which in
the continuum limit will typically be four dimensional, and have the 
graph structure of a tree.

In parts, this picture is consistent with the proposal of T. Hara and
G. Slade (mentioned in \cite{DS}) that the continuum limit of large 
clusters
looks like the process of Super Brownian Motion discussed in
\cite{Al, DS}).  However, unlike the compact SBM, we expect the 
scaling limit to exhibit clusters of arbitrary extent.  Our results 
suggest that in a certain 
(``weak'') sense the limit, which is still to be made sense of (!),
will exhibit two  features to which we are not 
accustomed in lattice percolation models:

\begin{itemize}
\item Percolation at the critical point.
\item Infinitely many infinite clusters.
\end{itemize}

The latter  may surprise one familiar with the general uniqueness 
Theorem  
of Burton and Keane \cite{BK}.  However, it should be appreciated that 
the BK result requires discreteness (and regularity on the 
corresponding scale) and thus is not applicable to the continuum limit.

\newpage

\startappendix 
\maappendix{The Relation between Proliferation 
of ISC and Hyperscaling}

The proliferation of the Incipient
Spanning Clusters is related to the breakdown of ``hyperscaling''.  
In order to clarify this relation, we recapitulate here  
the heuristic basis of the scaling and hyperscaling relations. 
Different variants of the argument can be found in the literature
\cite{Con,AGK,SA}.   The one given below is cast in terms
of quantities which are studied rigorously in this work.  However,
unlike in the rest of this paper, in this section we do not present 
rigorous results.  (Rigorous results on scaling relations
exist for $d=2$ \cite{K2} and $d>d_{u.c.}$ -- as mentioned in Sect.4)

The scaling relations of some of the critical exponents can be 
explained by the following picture.  
The first--line assumption, related to the 
self--similarity, is that the relevant quantities
scale by power laws.  Next:

\begin{enumerate}
	\item  For critical models, there are about $L^\#$ clusters in 
	$\Lambda_L$ with diameters of order $L$, and their volumes
	(defined on the U-V/lattice scale) are of the order of $L^D$.  
	If exceptions occur, it is assumed that they 
	do not affect the cluster statistics by more than 
	 $\times L^{o(1)}$. 
    The quantities we shall look at are
	\begin{equation}
	  \E\left( \sum_{\C\in\Lambda_L}
	      |\C|^k \ \I[\mbox{diam}(\C)\ge wL] \right)
		\label{A.1}
	\end{equation} 
	with ${\em w} < 1$, fixed as $L\to\infty$, and $k=1,2$. 
	  
	\item  For $p<p_c$ there is a characteristic length
	\be
       \xi(p) \ \approx \ (p_c-p)^{- \nu_{-}}
    \label{xi}
    \ee
   such that: 
   {\em i.\/} Only a fraction (say $1/3$, or 
   even up to $(1-L^{-o(1)})$)  
   of the mean value of the 
   cluster size $|\C(0)|$ is from distances greater than $\xi$.  \\  
   {\em ii.\/} Within the distance $\xi$, the two point function 
   $\tau_p(0,x)$ is comparable (in the sense of bounded ratios) with 
   its value at the critical point, $\tau_{p_c}(0,x)$.   
   
   Note that the mean cluster size and the two point function
   are in the direct  relation:
   \be
   \sum_x \tau_p(o,x) = \E_p(|\C(0)|)  \ \left( \approx 
   (p_{c}-p)^{-\gamma} \right) \  .
   \ee 

    \item  For $p>p_c$ the density of the infinite cluster, $M(p)$,  
    scales as
	  \be
	  M(p)\ \approx \ (p-p_c)^\beta 
	  \ee
	  and there is a characteristic length $\xi(p) \ 
	   \approx \ (p-p_c)^{-\nu_{+}}$
    such that $M(p)$ is of the order of the volume-fraction
    of $\Lambda_\xi$ which at $p=p_c$ belongs to clusters of 
    diameters comparable  
	with $\xi$ (i.e., Incipient Spanning Clusters on scale $\xi$).  
	
\end{enumerate}

\noindent Let us include in the list the assumption:
\begin{enumerate}
	\item[4.]    
	\be
	 \nu_{+}=\nu_{-}   \  ,
	 \label{nu}
	\ee
	\end{enumerate}
\noindent which is not essential for the picture 
presented here of hyperscaling, 
but seems to incorporate empirical observations, and fits well in the 
standard scaling ansatz. 

We shall compare now different ways of evaluating two quantities.  
First, let us look at
\be
\sum_{
	\begin{array}{l}
		 \C \mbox{ c.c. in } \Lambda_{L} \\
		\mbox{diam}\C \ge wL
	\end{array}   }  |\C|
\ = \ \sum_{x\in\Lambda} \I[\mbox{diam}\C(x) \ge wL]     \  .
\ee
Under the assumption {\em 3.} the mean value
of the expression on the right scales as $L^{d} L^{\beta/\nu_{+} }$.  
On the other hand, assuming {\em 1.}, the expression on the left
 scales as  $L^{\#}L^{D}$.  
Thus:
\be
 \mbox{\#}+D \ = \ d + \frac{\beta}{\nu_{+}}  
\label{5.7}
\ee

Next, consider 
\begin{equation}
	\sum_{\C \mbox{ c.c. in } \Lambda_{L}} |\C|^{2} \ =
	\ \sum_{x,y \in \Lambda_{L}} \I[\mbox{ $x$ and $y$ are 
	connected }] \ =
	\ \sum_{x\in \Lambda_{L}} |\C(x) \cap \Lambda | 
\end{equation}
Each of the expressions suggests a different method for evaluating
the 
mean value.  Using: {\em 1.} for the leftmost, the definition of 
$\eta$ (\eq{eta}) for the next, and the assumption {\em 2.} for the 
rightmost, one is led to: \be L^{\#}L^{2D}\ = \ L^{2d}/L^{d-2+\eta} \  
= \ L^{d}L^{\gamma/\nu_{-}} .  \ee i.e.,
\begin{equation}
	\mbox{\#} + 2D = d+2-\eta \ =\ d + \gamma/\nu_{-} \  .
\label{5.10}
\end{equation}

A convenient rearrangement of the information 
(equations (\ref{5.7}) and
(\ref{5.10}), incorporating {\em 4.}) is:
\bea 
  \fbox{ $ 2 - \eta = \gamma / \nu $ } & & \mbox{(scaling relation)} 
              \label{scaling} \\
	        \fbox{ $ D = \frac{\beta + \gamma}{\nu} $ }  &  &
	\mbox{(dimension of the  incipient spanning clusters)} \\
        	\fbox{ \# $= d - \frac{2\beta + \gamma}{\nu} $ }   & &
	        \mbox{(ISC proliferation exponent)} 
	          \label{hyperscaling relation}
	\eea
For independent 
percolation the proliferation exponent \# vanishes in 2D, 
and apparently also in all dimensions $d \le 6$. 
In such situations we get one more 
equation, which exhibits $d$ along the standard exponents,
and hence is called ``hyperscaling'':
\be
\fbox{ \# $=0$ , or $\ d - \frac{2\beta + \gamma}{\nu} = 0$ } 
\hspace{.3in} 
\mbox{(hyperscaling relation)} 
\ee
(That brings it to four equations for six quantities.)

Thus, the {\em breakdown of hyperscaling} is intimately related with 
the proliferation of the Incipient Spanning Clusters.  However, its 
validity does not require uniqueness --- just that the number of 
``macroscopic'' clusters be typically smaller than any power of $L$, 
or have a finite limit in a probabilistic sense.

\noindent {\bf Remarks:} {\em 1.\/} It is easy to 
incorporate in this picture  other 
 exponents, which were not listed above.  To make 
the list less incomplete, let us mention:
\be
\begin{array}{ccrl}
	\P\left( |\C(0)| \ge n \right) \approx n^{-(1/\delta)} & 
	\hspace{.5in} & \delta &= \frac{D\nu}{\beta}  \\
	 & & & \\ 
	  \frac{\mbox{card} \left\{ \C \subset \Lambda_{L} \ :\ 
      \mbox{diam } \C \ \ge\  wL  \right\} }{|\Lambda_L|} \ 
     \approx  \ (p_c-p)^{2-\alpha}    &   &  2-\alpha  
     & =   \nu(d-\mbox{\#}) \\
  &   &  & ( = 2 \beta + \gamma \ )   \  .
\end{array} 
\ee  
where we gave the definition 
(for  $\alpha$ not quite the standard one),
and a scaling relation derived along the above lines.

{\em 2.\/} The above discussion is relevant 
also for the Ising and Potts spin models, 
since the heuristic arguments 
described here make equal sense in the broader context of the
Fortuin -- Kasteleyn \cite{FK} random--cluster models.
The resulting scaling relations 
are the same in term of the recognizable characteristic exponents,
except that $d_{u.c.}$ is different and 
the significance of \# is lost if one is not aware of
the geometric structure behind the spin correlations.  (For Ising
spin systems, hyperscaling has also another connotation: 
its breakdown
implies that the scaling limit is 
a Gaussian random field \cite{A1,Bkr}).  
 
\newpage

\maappendix{Existence of Spanning Clusters in Critical Models}

In this Appendix, we prove Theorem 1, of Section 2.
A key role is played by the following estimate.  Both may be 
assumed to be known to experts.

\noindent {\em Claim:} 
In dimension $d>1$, for all $ s \le 1/3$:
\be
  Q_{L,s} \equiv \P\left(
	\begin{array}{l}
		\mbox{the set $[-sL,sL]^d$ is connected}  \\
		\mbox{to the boundary of $[-L,L]^d$}
	\end{array}  
	 \right) \ \ge \ C_d \ s^{(d-1)/2} \ .
\label{claim}
\ee
\noindent {\em Proof of the Claim:}  By monotonicity,
we may assume that $1/s$ is an integer. We shall show that if
\eq{claim} (with $C_{d}$ to 
be specified shortly) fails for some $L$ , then
the connectivity function decays exponentially (at distances $>> L$).  
That is well known to be in contradiction
with the condition $p=p_c$ \cite{Ham, AN}.   

To make the deduction, 
partition the lattice into cubic blocks of (linear) size $sL$.
For each self-avoiding path linking a site $x$ with $y$, 
let us associate a sequence 
of $sL$ blocks,  which are at distances approximately L apart, by the 
following algorithm.  
The zero-th block is the one containing $x$.  Next
is the block which the path reaches when it hits the boundary of 
the cube of size $(4+s)L$ concentric with the first block, and so on:
once a stopping point and a block are selected, we center on the $sL$ 
block a large cube of size $(4+s)L$, and
let the next stopping point be the exist site, and the next block be
the corresponding element of the lattice cubic partition.
Notice that for each block in this sequence, 
other than the end points,
the event seen in 
\eq{claim} occurs twice (at the end points once) and disjointly so.
The van den Berg - Kesten inequality \cite{vdBK}, which is described 
above, permits to deduce that the probability that 
$x$ and $y$ are connected by a self--avoiding path which 
corresponds to given sequence of $(k+1)$ blocks, 
is $\le Q_{L,s}^{2K}$.
Summing over the possibilities we get:
\bea
\tau(x,y) \ &\equiv 
& \ \P\left( \mbox{ $x$ and $y$ are connected } \right)
 \nonumber \\
\ & \le &  \sum_{k \ge |x-y|/(2L)}
 \left[ 2d\ 5^{d-1}\ \frac{1}{s^{d-1}} \ 
Q_{L,s}^{2} \right]^{K} 
\le  Const. \  e^{-\mu |x-y|} \  , 
\eea
with $\mu > 0$ if $\ 2d\ 5^{d-1}\ \frac{1}{s^{d-1}} \ 
Q_{L,s}^{2} < 1$. Since $\mu = 0$ 
at the critical point (\cite{Ham, AN}), 
we deduce that \eq{claim} holds, with $C_{d}= \sqrt{2d}\ 5^{(d-1)/2}$.

\noindent {\bf Proof of Theorem 1:}  If the probability that 
that the slab $S_{L,t}$ is traversed is very small, then so will be
 $Q_{L/2,s}$ for $s=1/2-t$.  To quantify that, 
let us consider two concentric cubes, $[-L/2,L/2]^{d}$ and 
$[-sL/2,sL/2]^{d}$.  If none of the 
$2d$ similar slabs
which enclose the smaller cube is spanned (in the short direction)
then that inner cube is not connected to the outer's boundary.
Since the $2d$ events are positively correlated (\cite{Har,FKG}), 
the probability of their simultaneous occurrence is greater than
the product.  Thus:
\begin{equation}
   \left[ 1 - Q_{L/2,s}  \right]^{2d}  \  \ge  \ 1 - R_{L,t} \  .  
	\label{RvssQ}
\end{equation}
The combination of \eq{RvssQ} with \eq{claim} shows that for 
$0<t<1/2$ the spanning probability is positive as claimed in 
\eq{2.1} with  $h(t) \ge C_{d}\ (1/2 - t)^{(d-1)/2} $.  
  
The crossing of narrow slabs, with $t\searrow 0$, can be estimated
by cutting the slab into $(3/t)^{d-1}$ smaller ones with the aspect
ratio close to $1/3$. {\em  Their} spanning probability is uniformly 
positive if $p \ge p_{c}$ (as shown by the previous discussion).  
Using the independence of the events, 
one obtains  \eq{2.2}.

\maappendix{Uniqueness for Supercritical Models}   \label{AppC}
 
In this appendix we supplement Theorem 2, by proving that 
only for $p=p_{c}$ would there be positive probability for observing
{\em more than one} spanning cluster in arbitrarily large systems.
We shall use the proven fact that in $d>2$ 
dimensions $p_{c}$ is the 
limit of the quadrant--percolation thresholds 
for slabs of finite width \cite{BGN,GM}.  This is an important 
technical statement, which  makes a variety of 2D arguments 
applicable to supercritical models in dimensions $d>2$
 \cite{ACCFR}.  (In particular, it permits to prove that the 
 spanning probability itself tends to $1$, as $L \to \infty$.)

\begin{thm}  For any $p\ne p_{c}$, and $t>0$,
\begin{equation}
   D_L(t,p) \equiv Prob_{p}\left(
	\begin{array}{l}
  \mbox{there is more than one spanning} \\
\mbox{cluster in $S_{L,t} ( \ \equiv[0,tL]\times[-L,L]^{(d-1)})$}
	\end{array}
	   \right)   \  \Ltoo \ 0
\label{app2.4}
\end{equation}
\label{thm6}
\end{thm}
\noindent {\bf Proof:}  For any $p<p_{c}$ the two point function 
decays exponentially \cite{AB,Men}, and that easily implies:
 $D_{L}(t,p) \le Const.  \ L^{2d} e^{-\mu L} \to 0$ (for $L\to 
 \infty$).  The assertion which still requires a proof is that 
 for $p > p_{c}$ (and $d>2$) the probability of there being more 
 than one  spanning cluster tends to $0$.

Let us define, for pairs of sites 
on the left boundary ($\{x_{1}=0\}$)
of the semi-infinite cylinder $[0,\infty)\times [0,L]^{d-1}$: 
\be 
G_{k}(x,y) = {\em Prob_{p}\/}\left(
\begin{array}{l}
	\mbox{ $x$ and $y$ are in distinct spanning clusters } \\
	\mbox{ of $[0,k]\times [0,L]^{d-1}$} 
\end{array}  \right)    \ .
\label{B.2}
\ee
The events seen in \eq{B.2} are obviously decreasing in $k$,
and thus the ratio \\
$G_{k+w}(x,y)/G_{k}(x,y)$ represents a 
conditional probability.   
For $p>p_{c}$, let $w(p)$ be the smallest slab width 
for  which there  is percolation in the quadrants 
$[0,w]^{d-2}\times [0,\infty)^{2}$  
($w(p) < \infty$  \cite{BGN, GM}).  
The above conditional probability, of the $(k+W)$-th event 
conditioned on the $k$-th event, is uniformly smaller that one, 
since there is a uniformly positive probability that the two 
distinct clusters will be joined by a path within the added slab.  
(The reason is
explained more explicitly in Section 4 of ref. \cite{ACCFR},
in the context of a rather similar argument.)  Thus
\begin{equation}
	G_{k}(x,y) \ \le G_{k-w}(x,y) \ e^{-\alpha} \ \ldots \ \le A \ 
	e^{-\alpha k/w}
\end{equation}
with some $\alpha > 0$.  Consequently:
\bea
	Prob_{p}\left( 
	\begin{array}{l}
		\mbox{there is more than one}  \\
		\mbox{spanning cluster in 	$\Lambda_{L}$}
	\end{array}	\right) 
	\  & \le & \sum_{
     \begin{array}{l}
		x, y \in \partial [0,2L]^{d}  \nonumber \\
		x_{1}, y_{1} = 0
	\end{array} }   G_{2L}(x,y) \\  \\ 
    & \le & \ A \ (2L)^{2d} e^{-\alpha 2L/w} \Ltoo 0 \  . 
    \nonumber
\eea

\remark We see that the probability of there being more than
one spanning cluster is exponentially small, in $L$,
for $p$ both above and below $p_{c}$,
 but not for $p=p_{c}$  (\eq{2.3}).  
This quantity yields a natural and meaningful characteristic length 
scale $\tilde{\xi}(p)$, and could perhaps offer a useful tool 
for a more thorough analysis of the the scaling relations.

\bigskip

\bigskip

\noindent {\large \bf Acknowledgments\/}
It is a pleasure to thank A. Aharony, J-P Hovi,  G. Slade, 
H.E. Stanley, D. Stauffer, and Yu Zhang for stimulating and 
enjoyable discussions of Spanning Clusters.  I wish to also express my 
gratitude for the hospitality enjoyed at 
the Department of Physics, Tel Aviv University, 
where some of the work was done.

\addcontentsline{toc}{section}{Acknowledgments and References}


\end{document}